\documentclass[sigplan,10pt]{acmart}
\usepackage{algorithm}
\usepackage[noend]{algpseudocode} 
\usepackage{mathtools}
\usepackage{tikz}
\newcommand*{\circled}[1]{\lower.7ex\hbox{\tikz\draw (0pt, 0pt)%
    circle (.5em) node {\makebox[.5em][c]{\small #1}};}}

\renewcommand\footnotetextcopyrightpermission[1]{} 
\usepackage{booktabs}   
\usepackage{subcaption} 
\usepackage{multirow}

\begin{document}

\title{Intelligent-Unrolling: Exploiting Regular Patterns in Irregular Applications}


\author{Changxi Liu}
\affiliation{%
	\institution{School of Computer Science and Engineering, Beihang University}
}
\email{changxi.liu@buaa.edu.cn}
\author{Hailong Yang}
\affiliation{%
	\institution{School of Computer Science and Engineering, Beihang University}
}
\email{hailong.yang@buaa.edu.cn}
\author{Xu Liu}
\affiliation{%
	\institution{Department of Computer Science, College of William and Mary}
}
\email{xl10@cs.wm.edu}
\author{Zhongzhi Luan}
\affiliation{%
	\institution{School of Computer Science and Engineering, Beihang University}
}
\email{07680@buaa.edu.cn}

\author{Depei Qian}
\affiliation{%
	\institution{School of Computer Science and Engineering, Beihang University}
}
\email{depeiq@buaa.edu.cn}



\begin{abstract}
\label{sec-abstract}

Modern optimizing compilers are able to exploit memory access or computation patterns to generate vectorization codes. However,  such patterns in irregular applications are unknown until runtime due to the input dependence. Thus, either compiler's static optimization or profile-guided optimization based on specific inputs cannot predict the patterns for any common input, which leads to suboptimal code generation. To address this challenge, we develop Intelligent-Unroll, a framework to automatically optimize irregular applications with vectorization. 
Intelligent-Unroll allows the users to depict the computation task using \textit{code seed} with the memory access and computation patterns represented in \textit{feature table} and \textit{information-code tree}, and generates highly efficient codes. Furthermore, Intelligent-Unroll employs several novel optimization techniques to optimize reduction operations and gather/scatter instructions. We evaluate Intelligent-Unroll with sparse matrix-vector multiplication (SpMV) and graph applications. Experimental results show that Intelligent-Unroll is able to generate more efficient vectorization codes compared to the state-of-the-art implementations.

\end{abstract}




\maketitle

\section{Introduction}
\label{sec-introduction}
With the SIMD instruction adopted on modern CPU architectures, the performance gap between CPU and memory become even larger. Compilers have developed powerful static analysis to accelerate applications automatically by leveraging the SIMD units on CPU. However, it works well only with regular applications. In addition, although the complex instructions such as reduction, gather and scatter have been supported on CPU architectures to optimize irregular applications, the performance with compiler optimizations is often sub-optimal. Especially when there are the potential write conflicts, the compilers usually give up on SIMD instructions trading performance for correctness. As the SIMD units become pervasive on modern CPU architectures, leaving the performance on table for irregular applications that take a large portion of scientific applications becomes unacceptable.

The regular applications can be optimized by static analysis of compilers for their memory access and instruction pattern. However, for irregular applications, the memory access and instruction pattern can only be analyzed during runtime. Therefore, the compilers fail to identify the performance opportunity for irregular applications. For instance, on the SIMD architecture, the compiler fails to optimize the program when confronting the potential write conflicts. However, if the runtime behavior of the data accesses can be identified, then we can solve the write conflicts for better parallelization using the SIMD units. Another example of compiler incapability at optimizing irregular applications can be found at the instruction level. For the gather/scatter/reduction instructions that are widely used in irregular applications, if we can organize the runtime data accesses are continuous or in the same vector lane, then we can replace the above instructions with load and permutation instructions for better performance.


However, there are several challenges to realize the potential performance opportunities for irregular applications that cannot be provided by compilers. Firstly, different from regular applications, the memory access and instruction pattern varies significantly across different irregular applications. Therefore, a general approach should be proposed to adapt to the various behaviors of irregular applications. Secondly, naively unrolling the instructions of irregular applications could lead to memory bloat that prevents further performance optimization. It is mandatory to constrain the memory occupancy when analyzing the runtime behaviors of irregular applications. Thirdly, the optimization method for irregular applications should be able to adapt to various underlying architectures in order to improve its practical adoption.


To address the above challenges, we propose Intelligent-Unroll, a framework for optimizing irregular applications on SIMD architectures automatically. There are three important components in Intelligent-Unroll, including code seed, feature table, and information-code tree.The design of Intelligent-Unroll is easily extensible by adding new features. Intelligent-Unroll have already integrated several optimization techniques for reduction, gather and scatter instructions for better performance. When evaluating with representative workloads, Intelligent-Unroll is able to generate more efficient codes on various SIMD architectures compared to the state-of-the-art implementations. 


Specifically, this paper makes the following contributions:
\begin{itemize}
	\item We propose Intelligent-Unroll, a framework that identifies the regular patterns within irregular applications, and automatically optimize the instruction and data synthetically by generating more efficient codes.
	\item We propose several techniques such as code seed, feature table and information-code tree to identify the opportunity to replace the reduction instructions with load instructions, and the gather instructions with instruction group of load, shuffle and select instructions for better performance.
	\item We evaluate with representative workloads such as SpMV and PageRank on KNL and Intel Xeon CPUs. The experiment results demonstrate that the codes automatically generated by Intelligent-Unroll achieve better performance than the state-of-the-art implementations.
\end{itemize}

The remainder of this paper is organized as follows. Section \ref{sec-background} presents the background of the irregular application and corresponding optimization methods. Section \ref{sec-motivation} describes the motivation of our work. Section \ref{sec-overview} presents the design overview of \textit{Intelligent-Unroll}. Section \ref{sec-reduction} and Section \ref{sec-gather-scatter} describes the implementation details of the optimizations on reduction and gather operators. Section~\ref{sec-evaluation} presents the evaluation results of SpMV and PageRank compared to the state-of-the-art implementations. Section \ref{sec-relatedWork} presents the related work in the field, and section \ref{sec-conclusion} concludes this paper.

\section{Background}
\label{sec-background}
\subsection{Understanding Irregular Applications}
Irregular applications are common in both traditional research fields such as high performance computing and emerging research fields such as big data analysis and deep learning, which exhibits a constant demand for higher performance. The difference between irregular and regular applications is that whether the patterns of data access and instruction can be known before runtime. For irregular applications, the above patterns is strongly correlated with the input data and can only be known during runtime. Such uncertainty of irregular applications introduces difficulties such as irregular memory accesses, unbiased branches and writing conflicts for compiler optimization. 

For irregular applications, there are two important concepts to describe their data access and instruction patterns such as \textit{access arrays} and \textit{data arrays}~\cite{ding1999improving}. Algorithm~\ref{alg-irregular} presents two code example of irregular applications. We can see that the \textit{access arrays} contain the indirect access or branch execution sequence (line~\ref{alg-irregular-index} and line~\ref{alg-irregular-branch}). Whereas the \textit{data arrays} are almost accessed indirectly through the \textit{access arrays} (line~\ref{alg-irregular-data}).
Another code example of irregular applications is the inference process of the sparse neural networks~\cite{Liu_2015_CVPR,park2016faster}, although the data arrays during the inference can be updated, the access arrays are immutable or updated infrequently. The above observations inspire us to design a mechanism for uncovering the potential performance of irregular applications and applying corresponding optimization automatically. 

\begin{algorithm}[hbtp]
	\caption{The code examples of irregular applications}
	\label{alg-irregular}
	\begin{algorithmic}[1] \footnotesize
		\Function{Irregular Memory Access}{}
		\State $idx \gets Load\ access\_array[...]$\label{alg-irregular-index}
		\State $data \gets Load\ data\_array[idx]$\label{alg-irregular-data}
		\EndFunction
		\Function{Unbiased Branches}{}
		\State $cond \gets Load\ access\_array[...]$
		\If { $cond$ }\label{alg-irregular-branch}
			\State $...$
		\EndIf
		\EndFunction
	\end{algorithmic}
\end{algorithm}

\subsection{Optimizing Irregular Applications}
The performance gap between CPU and memory is still increasing. Although multi-level memory hierarchies are introduced to hide memory access latency, it still cannot catch up with the instruction level parallelism developed in hardware such as SIMD, multi-stage pipeline and out-of-order execution. For regular applications, the compilers can generate efficient instructions such as AVX512 through static analysis of the program patterns for optimized performance. However, with irregular applications, the compiler optimization is quite restricted due to the unknown data access and instruction patterns that can only be determined during runtime. For instance on the SIMD architectures, to ensure the correctness, the compilers perform almost no vectorization of irregular applications if there are potential memory write conflicts. The conservative optimization strategy of existing compilers wastes the opportunities to exploit the regular patterns within irregular applications for performance optimization. 

Similar to regular applications, the optimization of irregular applications also focus on the temporal and spatial reuse of data, as well as parallel efficiency. There are plenty of research works proposed to adapt irregular applications to underlying architectures. However, most of the above works require tremendous engineering efforts and cannot be easily ported to other architectures. Such ad-hoc optimizations are unsustainable as new architectures and applications are developed at unprecedented rate especially in the emerging domains such as deep learning. In addition, the optimization of irregular applications has also been studied in domain specific compilers such as Halide~\cite{ragan2013halide}, Tensor Comprehensions~\cite{vasilache2018tensor} and TVM~\cite{chen2018tvm,devito2011liszt}. These studies provide efficient way to generate high performance code for special application domains. These domain specific compilers motivate our work to design a compilation framework for irregular applications that can analyze the data access and instruction patterns to generate efficient code automatically. We choose LLVM~\cite{lattner2004llvm} as our compilation backend, because the JIT APIs in LLVM allow us to analyze the execution patterns and generate optimized code at runtime.
\section{Motivation}
\label{sec-motivation}
The memory access patterns of regular applications have already been optimized by the compilers using static analysis~\cite{nuzman2006auto,anderson2016automatic}. However, the memory patterns of irregular applications are always dictated by the data being processed, which can only be known during runtime. Therefore, the existing compilers fail to optimize the performance for such irregular applications.

\begin{figure}
	\centering
	\includegraphics[scale=0.28]{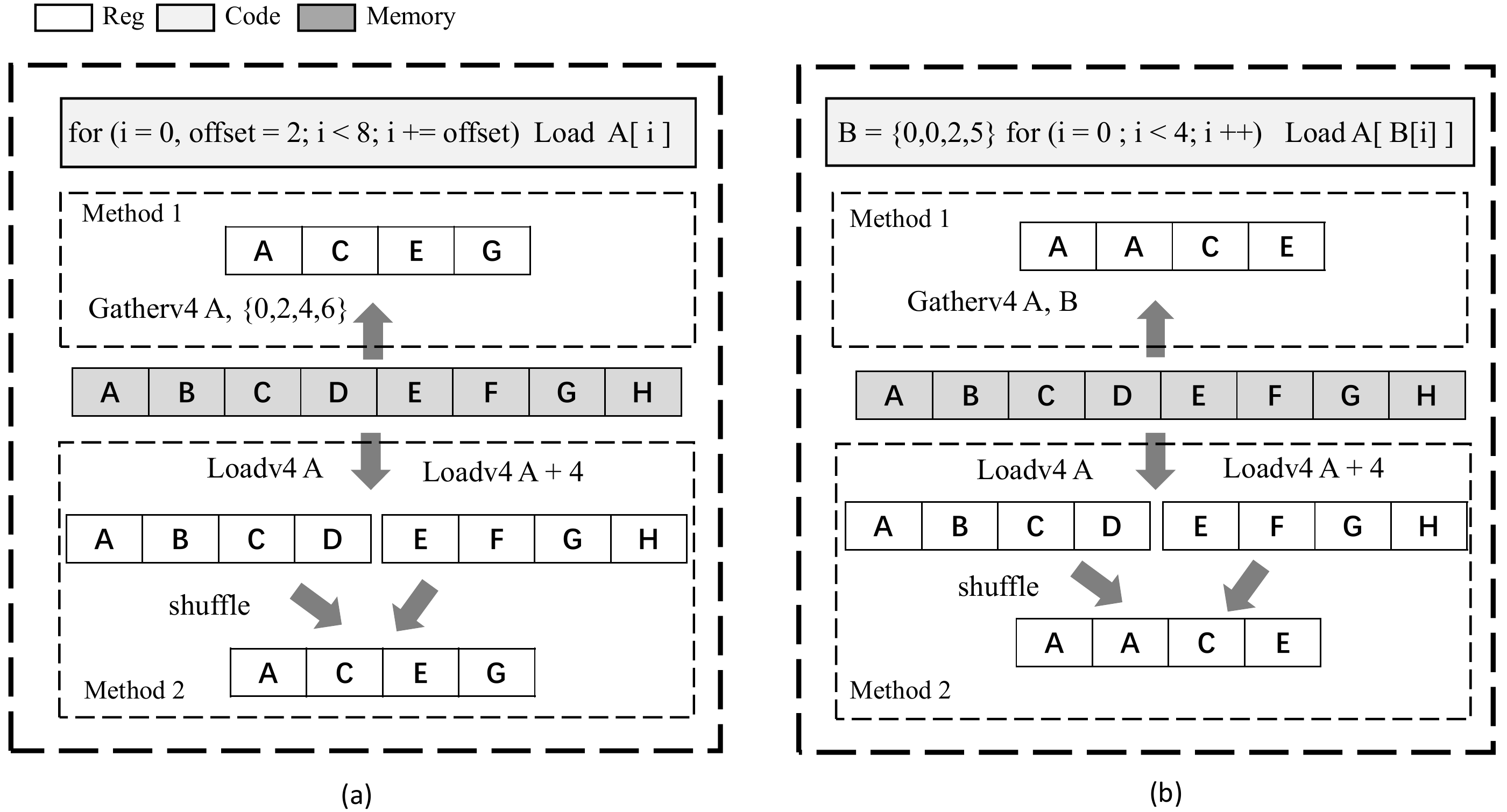}
	\caption{The memory access optimization of regular application (a), and irregular application (b).}
	\label{fig-motivation-gather}
\end{figure}

The memory access pattern usually has a significant impact on the performance. However, using the existing compiler optimizations sometimes lead to suboptimal performance for irregular applications. For instance, when loading the data from discontinuous memory addresses, the compilers alway generate \textit{gather} instruction for the memory load. However, as shown in the case of Figure~\ref{fig-motivation-gather}, replacing the \textit{gather} instruction (Method 1) with \textit{vload} instruction (Method 2) achieves better performance. In the case of the regular application as shown in Figure~\ref{fig-motivation-gather}(a), the compilers can automatically perform the above optimization through static analysis. Whereas with irregular applications as shown in Figure~\ref{fig-motivation-gather}(b), since the memory access pattern can only be recognized during runtime, the compilers generate inefficient codes that load data from memory using \textit{gather} instruction.
 
\begin{figure}
	\centering
	\includegraphics[scale=0.28]{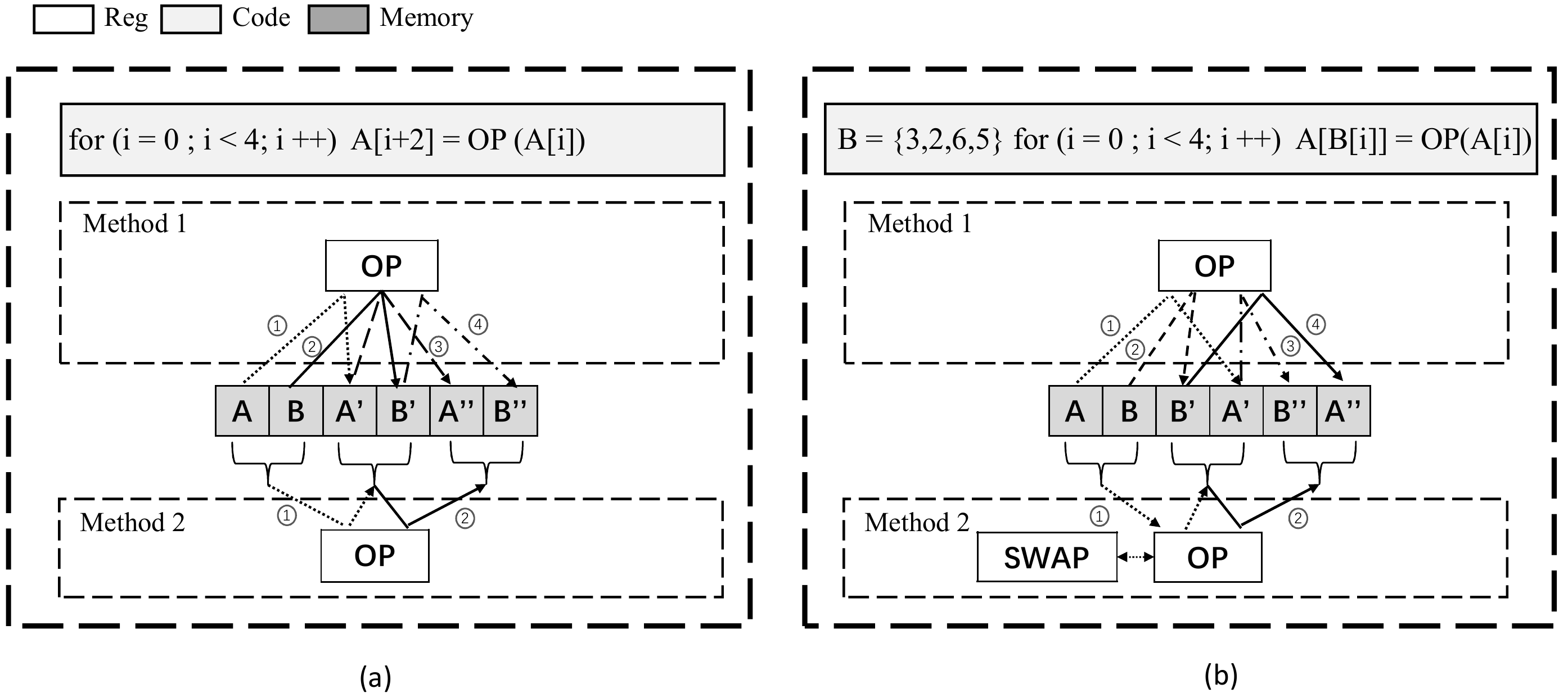}
	\caption{The calculation optimization of regular application (a), and irregular application (b).}
	\label{fig-motivation-cal}
\end{figure}

Moreover, existing compilers are incapable to generate efficient code for the calculation of irregular applications. For instance, to utilize the vector units on SIMD architectures, the calculation dependencies need to be identified for correct vectorization. For regular application as shown in Figure~\ref{fig-motivation-cal}(a), compilers can identify the calculation dependencies with static analysis and then generate efficient code. For instance, the operation\circled{1}, operation\circled{2}, operation\circled{3} and operation\circled{4} are independent from each other. Then, the compilers can leverage such information (Method 2) to optimize performance. However, when dealing with the irregular application in Figure~\ref{fig-motivation-cal}(b), the compilers have to assume that the calculations have dependencies with each other to ensure correctness. Whereas the optimization (Method 2) in Figure~\ref{fig-motivation-cal}(b) indicates that operation\circled{1} and operation\circled{2} can be processed in parallel, and then operation\circled{3} and operation\circled{4} can be processed in parallel, which leads to better performance. Unfortunately, such optimization opportunity of irregular applications cannot be identified by compilers using static analysis.

The above observations indicate that there is a huge space for performance optimization of irregular applications that cannot be achieved by compilers using static analysis. Such performance opportunity within irregular applications can only be identified during runtime that involves both memory accesses and calculation instructions.

However, naively unrolling the instructions of irregular applications and then applying optimizations could easily generate formidable code space, that leads to the instruction bloat problem. In addition, if we use condition statements to select the optimal instructions, the application performance could degrade significantly due to the branch mis-prediction caused by the condition statements. Moreover, empirically writing specific code for each condition is also impractical, which requires tremendous engineering efforts. For instance, If the conditions to be optimized are ($k1,k2,k3...$), then the number of code to be written is ($k1 \times k2 \times k3 \times...$).

To overcome the above problems, we propose Intelligent-Unroll, a framework that allows users to provide a code seed to describe the calculation process of the program. Intelligent-Unroll then automatically generates efficient instructions for the program. Specifically, Intelligent-Unroll can identify the regular instruction patterns and optimize them with efficient instructions. To accomplish above goals, Intelligent-Unroll provides corresponding techniques to tackle the following challenges:

\begin{itemize}
	\item How to leverage the code seed to describe diverse data access and instruction patterns? 
	\item How to adapt instructions to the behaviors of data accesses for better performance? 
	\item How to optimize the instruction and data access synergistically without violating the correctness?
\end{itemize}

\section{Intelligent-Unroll: Overview}
\label{sec-overview}
\begin{figure*}
	\centering
	\includegraphics[scale=0.64]{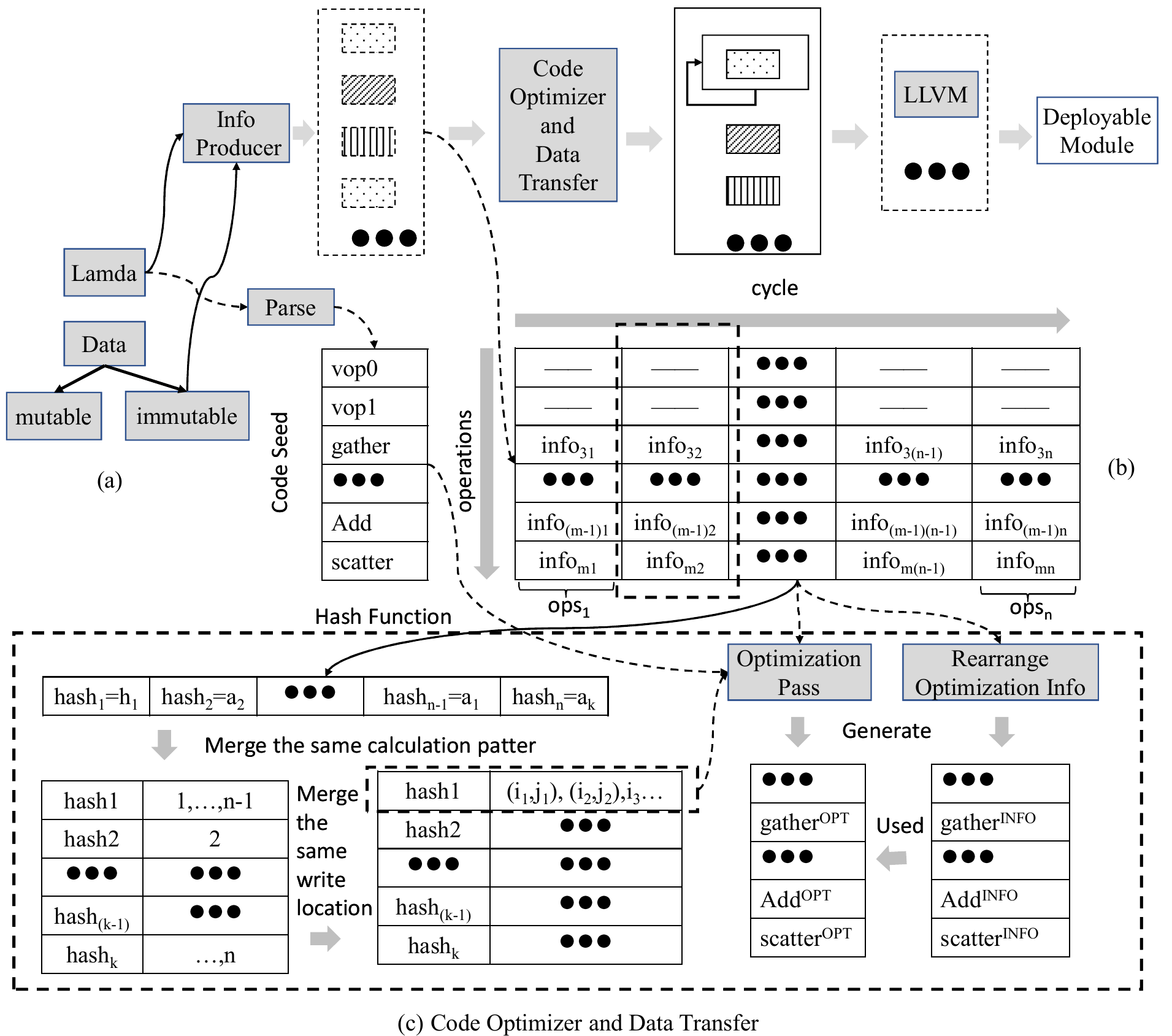}
	\caption{The design overview of \textit{Intelligent-Unroll}, which includes (a) information producer, (b) feature table and (c) code optimizer and data transfer.}
	\label{fig-overview}
\end{figure*}
\begin{figure}
	\centering
	\includegraphics[scale=0.35]{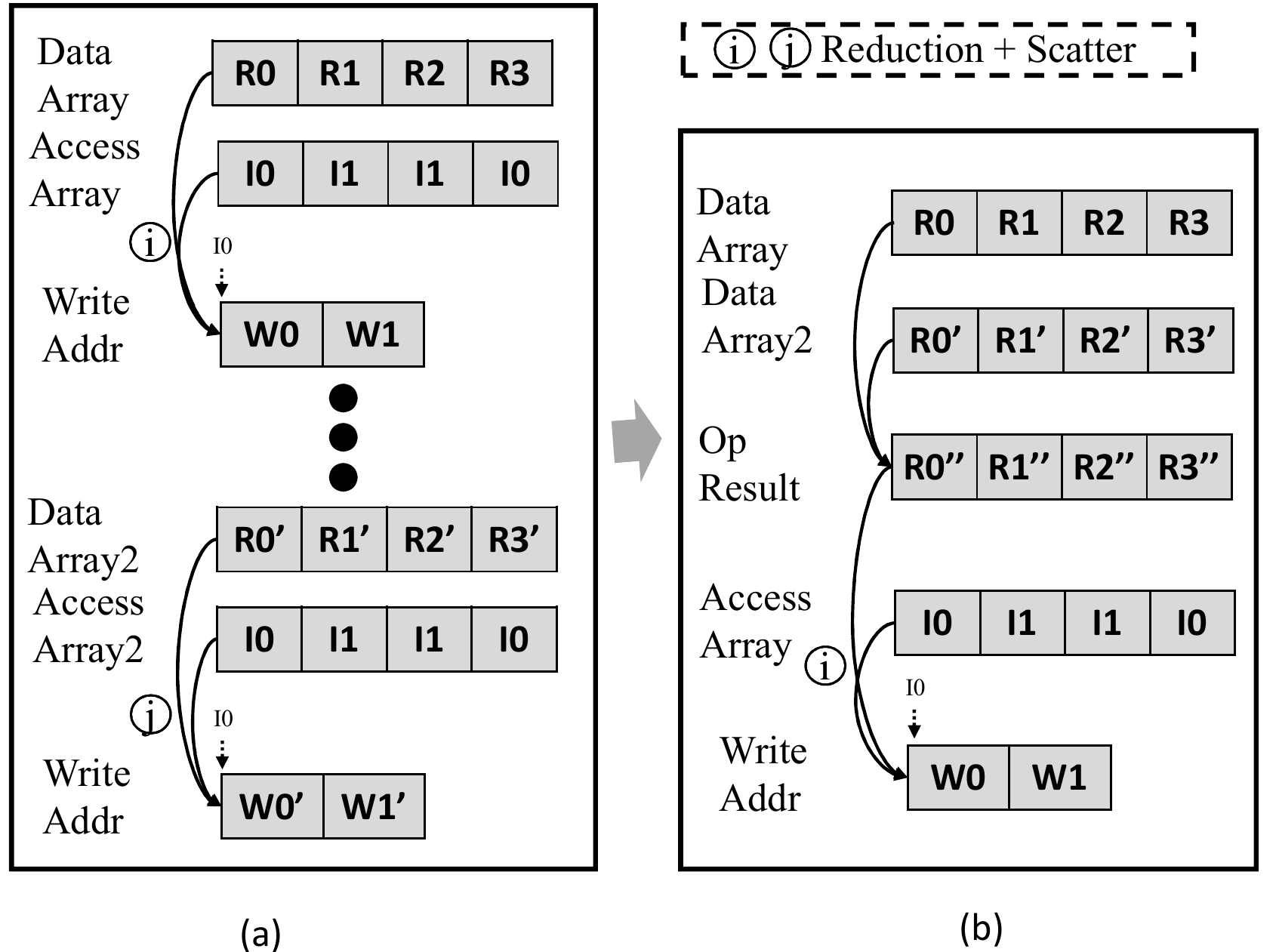}
	\caption{An example of merging same location instruction groups together (a) the instruction merged before. (b) the instrcution merged after}
	\label{fig-merge-write}
\end{figure}
Intelligent-Unroll is designed to identify the regular data access and instruction patterns hidden deeply within irregular applications. The goal of Intelligent-Unroll is to automatically optimize the instruction and data synthetically for identified performance opportunities.

The design overview of Intelligent-Unroll is shown in Figure~\ref{fig-overview}. The users only need to describe the calculation process using a lambda expression with its input data, and then Intelligent-Unroll interprets calculation expression and automatically generates an efficient implementation for a particular architecture. The data of the computation task is classified to mutable data and immutable data. The immutable data, that is unchanged during the execution of the task, will be used to generate information for the optimization process. For the optimization process, Intelligent-Unroll firstly interprets the lambda expression and generates the code seed. The instruction patterns contained in the code seed as well as the immutable data are used by the \textbf{Information Producer} (Figure~\ref{fig-overview} (a)) to generate the \textbf{Feature Table} (Figure~\ref{fig-overview} (b)), which includes the information required for further optimization. 

The \textbf{Code Seed} describes the calculation process without concerning about the optimization. Based on the \textit{Code Seed}, the \textit{Information Producer} extract the calculation patterns to generate the \textit{Feature Table}. And \textbf{Code Optimizer and Data Transfer} modules use the \textit{Code Seed} to generate optimized code. Each column of the \textit{Feature Table} is the calculation process for one iteration, and the row represents the iterations. Each element in the \textit{Feature Table} describes the instruction feature at the current iteration. Each column of the \textit{Feature Table} is denoted as $ops_k$, where $k$ is the $k$-th order. The \textit{Feature Table} helps us handle various patterns in the irregular applications. We can merge instructions to optimize the execution based on the information provided by \textit{Feature Table}.

The \textit{Code Optimizer and Data Transfer} modules in the \textit{Information Producer} then process the \textit{Feature Table} to generate the Intermediate Representation (IR) code that is independent from the underlying architecture. Eventually, Intelligent-Unroll lowers the the code implementation to LLVM to generate the machine instruction regarding the target architecture.

The design of \textit{Code Optimizer and Data Transfer} modules is shown in Figure~\ref{fig-overview}(c). Firstly, the hash value of each column in the \textit{Feature Table} is generated. The columns with the same hash value exhibits the same calculation pattern. Intelligent-Unroll merges the columns with the same hash value to generate a hash map. This hash map combines the instructions with the same calculation pattern, and thus deceases the memory occupancy during instruction unrolling. 

After combining instructions, the Intelligent-Unroll continues to process the hash map to merge instructions with the same write location. Figure~\ref{fig-merge-write}(a) shows an example of two instruction groups writing to the same location. Without merging the instructions, two reduction operations in addition to two read and write operations to the \textit{Write Addr} are required, which wastes computation resources and memory bandwidth. Figure~\ref{fig-merge-write}(b) shows the calculation pattern after merging the instructions. We can see that only one reduction operation is required. Although in this case we introduce one extra vector operation, it is far more efficient than reduction operation. Eventually, the optimized instructions are generated by the \textbf{Optimization Pass} and \textbf{Rearrange Optimization Info} modules, the details of which are described in Section~\ref{sec-reduction} and Section~\ref{sec-gather-scatter}.

\section{Reduction Instruction}
\label{sec-reduction}
The reduction instruction is a frequently used in programs. However, the reduction instruction encounters the instruction dependency problem on the SIMD architecture for parallelization. Traditional compilers degrades to SISD instructions because it fails to identify the dependency using static analysis. The pseudo-code shown in Figure~\ref{fig-reduction} serves as an example. However, naively applying vectorization could lead to incorrect results, for example more than two operators writing to the same location in one SIMD instruction.

In Intelligent-Unroll, it can analyze the write locations and rearrange the calculation to avoid write conflicts. However, changing the original calculation order may jeopardize the correctness of the program, therefore we need to make sure the correctness is not affected by the calculation rearrangement. The analysis of the calculation rearrangement in terms of program correctness is as follows.

The reduction operator is both associative and commutative. We define the reduction operator as $*$, and thus an example of reduction operation can be expressed as $res = p1 * p2 * p3 * p4$. The expression can be transformed to $res = (p1 * p2) * (p3*p4)$ based on the associative property. Therefore, we can calculate $res1 = p1 * p2$ and $res2 = p3 * p4$ in parallel, and then calculate $res = res1 * res2$. It is clear for the reduction operation that we reduce the partial results in parallel and then reduce partial results to derive the final results.

\subsection{Generating Information}
Instead of generating code by the distribution of write locations, we generate various reduction operators by the number of reduction operations required. On the SIMD architecture whose vector length is $N$, we need $log(N)$ reduction instructions at most to complete a SIMD reduction operator. We denote a flag of the reduction operator, which ranges $0,1,2,...,log(N)$. For example, when the flag of the reduction operator is $M$, it means that we need $M$ reduction instructions to complete the  SIMD reduction operator.

\begin{figure}
	\centering
	\includegraphics[scale=0.35]{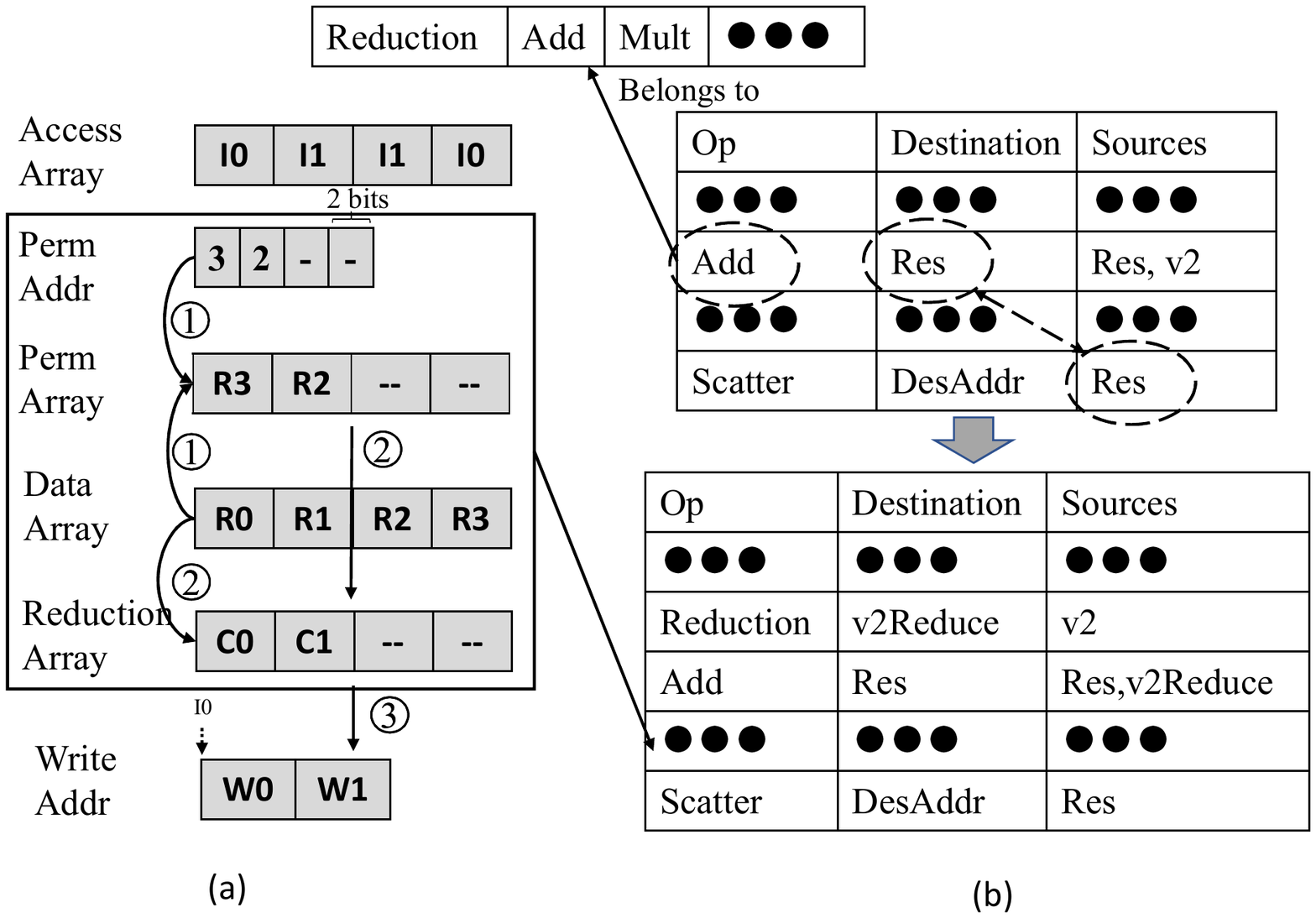}
	\caption{An example of reduction operator (a) and, (b) corresponding code generation pattern.}
	\label{fig-reduction}
\end{figure}

In addition to the flag, we also need other information. When the flag is $M$, it requires M vector, whose dimension is $N$ and the bit width of each element is $log(N)$. The above information represents the source location of the data to be reduced. As shown in Figure~\ref{fig-reduction}(a), R3 requires a reduction operator with R0, R1 and R2 each. Therefore, the shuffle address is 3 and 2, and R3 and R2 are moved to the first and second location of the shuffle data. We can reduce the shuffle data and then the rest of the data together to derive the final results. When the flag is $N$, we can also choose the reduction operator supported by the architecture if it is available.


\subsection{Identifying Code Generation Pattern}
The commonly used reduction operators include add and multiply. For other reduction operators such as minus, division, we can transform them to add or multiply reduction operators with negative variance operators.

The code seed generated does not consider the write conflicts and the optimization pass module after will process it. Intelligent-Unroll identifies the source instruction that provides the write variance of scatter instructions. The reduction processing module is activated to insert several reduction operations before scatter instructions, if the operation type of the source instruction belongs to the reduction operators. Intelligent-Unroll will generate reduction instructions according to the information in the column of \textit{Feature Table} corresponding.

As shown in Figure~\ref{fig-reduction}(b), the \textit{Res}, which is the value written by \textit{Scatter} instruction, is provided by an \textit{Add} operation, which belongs to reduction operators. Activated by this condition, Intelligent-Unroll inserts a reduction operation before the \textit{Add} instruction and then redirects the result to the \textit{Add} instruction, which is the operation 1 and 2 in the Figure~\ref{fig-reduction}.

\subsection{Instruction and Memory Efficiency}
\begin{table}

	\centering
	\scriptsize
	\caption{The comparison of the instructions before and after the optimization of reduction operator.}
	\begin{tabular}{c|c|c|c|c|c}
		\hline
		  &Calculation & Reduction  & Permulation \\
		\hline
		
		original & N  & N & 0 \\ \hline
		optimized &  1  & M & M \\ \hline
	\end{tabular} \label{tab-reduction-cal}
\end{table}
\begin{table}
	
	\centering
	\scriptsize
	\caption{The comparison of the data size before and after the optimization of reduction operator.}
	\begin{tabular}{c|c|c|c|c}
		 \hline
		 & \multicolumn{3}{c|}{vload}                  & vstore\\ \hline
		
		 & Write Index&Write Data &Additional Data& Write Data \\ \hline
		
		original   & N * Bit(Index) & N * Bit(Data) & --            & N * Bit(Data) \\ \hline
		optimized    & M * Bit(Index) & M * Bit(Data) & M * Bit(Info) & M * Bit(Data) \\ \hline
	\end{tabular} \label{tab-reduction-mem}
\end{table}
Intelligent-Unroll generates optimized codes for the original program. Table~\ref{tab-reduction-cal} provides a comparison of the instructions before and after the optimization. With Intelligent-Unroll, we can reduce the number of calculations on the reduction data from N to 1, and the number of reduction operations from N to M, where M is less than or equal to $log_{2}{N}$. Although Intelligent-Unroll introduces additional operations such as \textit{Permulation}, it can still accelerate the calculation process if executing M shuffle operations is faster than the sum of (N-1) calculations and (N-M) reduction operations.

Intelligent-Unroll also changes the memory access pattern. From Table~\ref{tab-reduction-mem} we can see, it avoids the redundant memory load and store to the write data, whose size is $(N-M) \times Bit(Data)$. In addition, Intelligent-Unroll also eliminates unnecessary load to the index of write address, whose size is $(N-M) \times Bit(Index)$. However, Intelligent-Unroll also introduces extra overhead. The additional data that is used by the shuffle instructions is $M \times log_{2}{N}bits$. Therefore, the performance of memory access can be optimized if the size of additional data is less than the sum of the write data size after optimization.

\section{Gather and Scatter Instructions}
\label{sec-gather-scatter}
\begin{figure}
	\centering
	\includegraphics[scale=0.40]{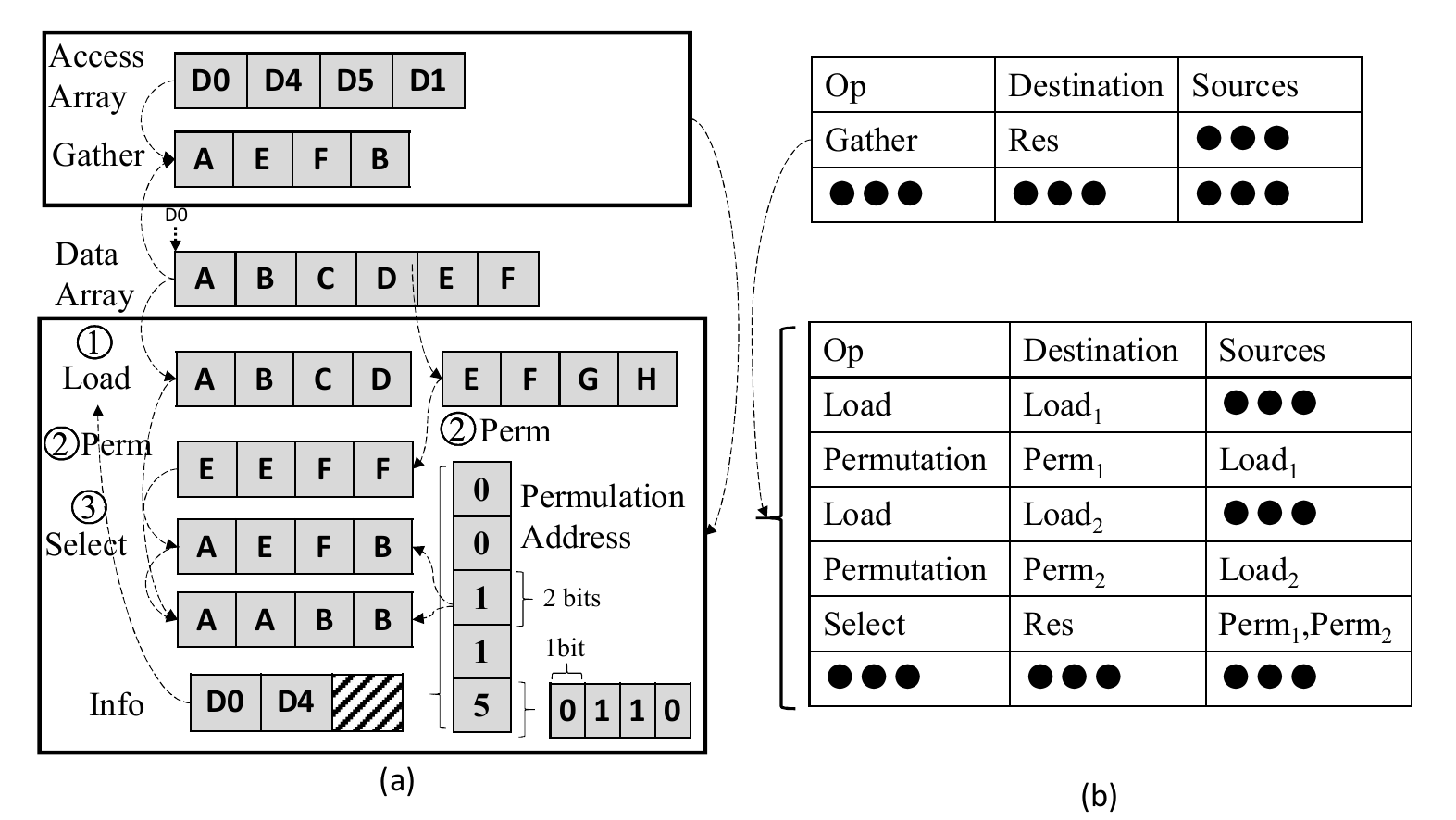}
	\caption{An example of gather operator (a) and, (b) corresponding code generation pattern.}
	\label{fig-gather}
\end{figure}

\subsection{Understanding the Opportunity}
Gather and Scatter instructions are also frequently used in programs on SIMD architectures. We observe that replacing the gather instruction with group of \textit{vload} and \textit{permutation} instructions achieves better performance in several cases. Similar performance improvement is also observed by replacing \textit{scatter} instruction with group of \textit{permutation} and \textit{store} instructions. Since the method of optimizing \textit{gather} and \textit{scatter} instructions is similar, we only present the optimization method of \textit{gather} instruction in the following.
\begin{figure}
	\centering
	\includegraphics[scale=0.28]{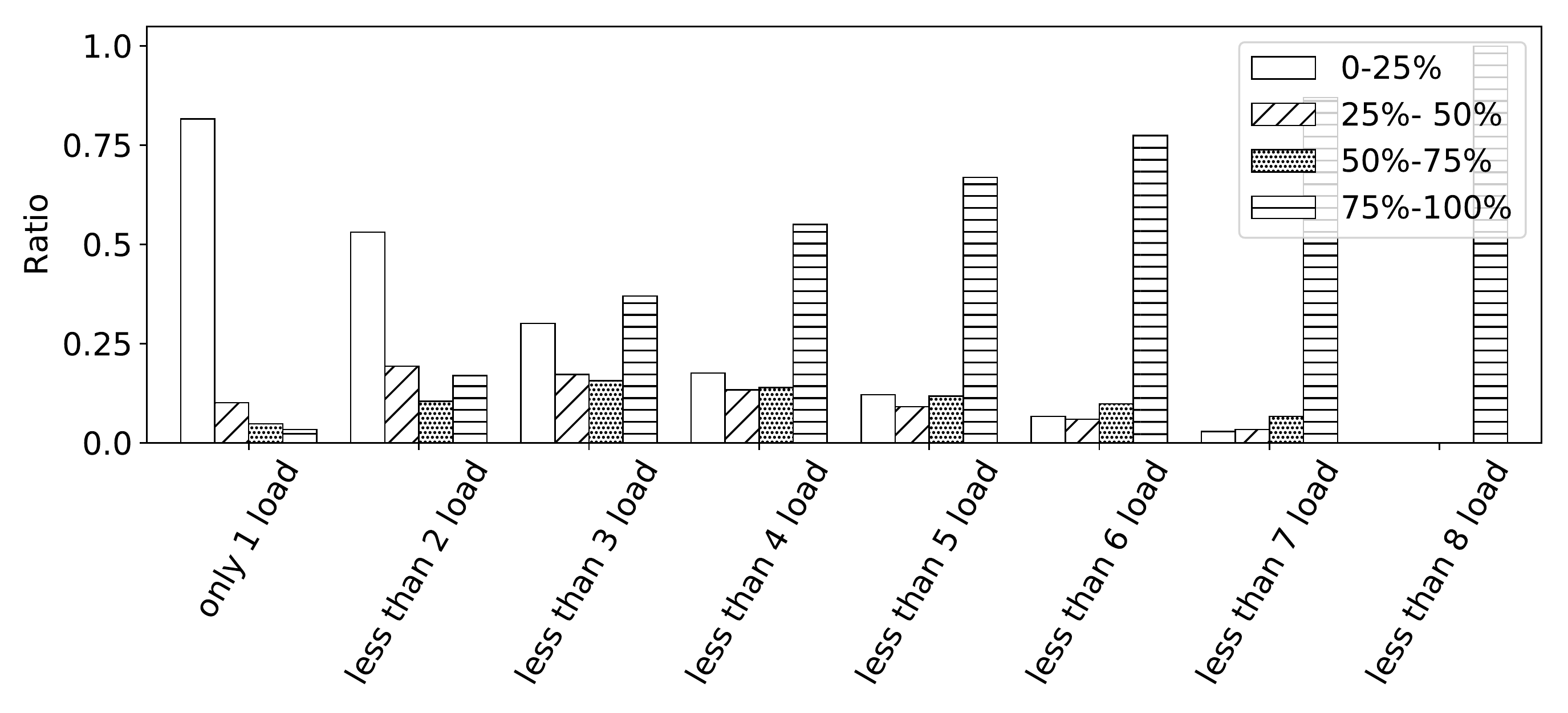}
	\caption{The distribution of gather instructions that can be replaced by instruction group of vload, permulation.}
	\label{fig-gather-distribute}
\end{figure}

Unlike the reduction instruction, the sparsity pattern of the data affects the performance opportunity when optimizing the gather operator. For instance, if the sparsity of the data is entirely random, there is hardly a chance to achieve better performance. Fortunately, most of the sparse data exhibit regular distribution to some extent. Figure~\ref{fig-gather-distribute} shows the percentage of sparse datasets that achieve better performance when replacing the gather instructions with vload instructions. The sparse datasets include 2,700 matrices from the SuiteSparse Matrix Collection~\cite{davis2011university}. The x axis in the figure  indicates the number of vload instructions, and the y axis indicates the percentage of the entire datasets. The legend of the Figure~\ref{fig-gather-distribute} represents the percentage of the gather instructions within the execution on a particular dataset.


From Figure~\ref{fig-gather-distribute} we can see that the datasets, with more than 25\% of the gather instructions can be replaced by one vload instruction, accounts for 18.4\% of the entire datasets. Whereas, 46.9\% of the datasets contain more than 25\% of the gather instructions that can be replaced with no more than two vload instructions. Moreover, 55.0\% of datasets contain more than 75\% of the gather instructions that can be replaced with four vload instructions. It is clear that there is a large performance space by optimizing the gather instructions of irregular applications with sparse data.

\subsection{Generating Information}
Similar to the optimization of reduction operator, we use a flag to denote the number of vload instructions, and the largest value of the flag is vector length of the architecture. 
And the same to the reduction operators, the optimization of \textit{gather} instructions also need additional information and the bit width of each element in the  address vector and the length of the vector is the same as reduction operator. The difference from reduction operator is that we use only one \textit{Permulation Address} regardless the value of the flag. To determine the \textit{permulation} instruction that data in the address vector belongs to, we use additional mask vector whose number is ($flag - 1$). Several begin addresses are also required whose value equals to the flag in order to guide the vload instructions.

The Figure~\ref{fig-gather} gives an example of optimizing \textit{gather} instructions. The Figure~\ref{fig-gather}(a) is an example of gather operator, where the vector length is four, the bit width of the shuffle vector is two, and the length of vector is four. In this example, we use two vload instructions to replace one gather instruction. Therefore, the value of the flag is two, and the number of the mask vector is one. First, we load data ABDC and EFGH in the registers using the begin addresses D0 and D4. Then, based on the \textit{Permutation Address} and ABCD,EFGH, we obtain AABB, EEFF by permutation instruction. After that, we obtain AEFB with AABB, EEFF with mask 0110 using the \textit{select} instruction.

\subsection{Identifying Code Generation Pattern}
To optimize the \textit{gather} instructions, we replace the \textit{gather} instructions with \textit{vload}, \textit{permutation} and \textit{select} instructions. When scanning the code, we consult the column of feature table corresponding to determine whether there is performance benefit by replacing the \textit{gather} instruction with the instruction group (e.g., \textit{vload}, \textit{permutation} and \textit{select} ). Then, Intelligent-Unroll performs the code transformation to generate the optimized code. Figure~\ref{fig-gather}(b) shows an example of the code generation for gather operator. The instructions including multiple \textit{vload}, \textit{permutation} and \textit{select} instructions is used to replace the original \textit{gather} instructions. And if the flag value equals to one, it only requires \textit{vload} and \textit{permutation} instructions.
\subsection{Memory and Efficiency}
\begin{table}
	\centering
	\scriptsize
	\caption{The comparison of the data size before and after the optimization of gather operator.}
	\begin{tabular}{c|c|c|c|c}
		\hline
		
				 &Index			  & Data 		  & Additional Info \\ \hline
		
		original   & N * Bit(Index) & N * Bit(Data) & --              \\ \hline
		optimized    & M * Bit(Index) & M * N * Bit(Data) & $ N * log_{2}{N} + (M - 1) * N $   \\ \hline
	\end{tabular} \label{tab-gather}
\end{table}
As shown in Table~\ref{tab-gather}, after our optimization of \textit{gather} operator, the number of index data avoided to be loaded is $N-M$. However, our optimization introduces $(M-1) \times N$ extra data to be loaded as well as $ N \times log_{2}{N} + (M - 1) \times N $ bits to record the additional information. In addition to the memory load overhead, our optimization also requires $M$ instruction groups of vload, permutation and select instructions. 

Fortunately, on the cache hierarchy of modern processor, the number of cache lines consumed by our method is the same as the original \textit{gather} instruction. In addition, the size of the extra data introduced by our method  is always smaller than the size of index data eliminated. Since our method is effective when the performance improvement with the optimized gather operator outweighs the overhead due to the extra data, we apply the optimization only when the flags indicate there are performance benefits. 

\section{Evaluation}
\label{sec-evaluation}
\subsection{Experiment Setup}
\begin{algorithm}[t]
	\caption{The code snippet of SpMV in CSR format}
	\label{alg-spmv}
	\begin{algorithmic}[1] \footnotesize
		\For{$i \gets 0 , m$}
			\For {$j \gets row\_ptr[i], row\_ptr[i+1]$}
				\State $y[i] \gets y[i] + value[j]\times x[col\_ptr[j]]$
			\EndFor
		\EndFor
	\end{algorithmic}
\end{algorithm}
\begin{algorithm}[t]
	\caption{The code snippet of PageRank}
	\label{alg-pagerank}
	\begin{algorithmic}[1] \footnotesize
		\For{$i \gets 0 , nedges$}
			\State $sum[n2[j]] \gets sum[n1[j]] + rank[n1[j]]\  /\  nneighbot[n1[j]]$
		\EndFor
	\end{algorithmic}
\end{algorithm}
\newcommand{\tabincell}[2]{\begin{tabular}{@{}#1@{}}#2\end{tabular}}
\begin{table*}
	\centering
	\scriptsize
	\caption{The platform and benchmark evaluation approach. All experiments are done with single thread.}
	\begin{tabular}{|c|c|c|}
		\hline
		The platform & SpMV evaluation approach&PageRank evaluation approach\\ \hline
		\tabincell{l}
		{\textbf{Intel Phi 7210}
		 \\(64 cores@1.3GHz,2.66 DP TFlops,\\
		 16GB MCDRAM,400GB/s bandwidth,\\
		 384GB DDR4,102.4Gbit/s bandwidth).}	 
	     &\tabincell{l}{
		 (1) The CSR-based SpMV compiled by ICC.\\
	 	 (2) The CSR-based SpMV with Intel MKL version 2019 Update 3.\\
 	 	 (3) CSR5-base SpMV\cite{liu2015csr5}.\\
  	 	 (4) The code generated by Intelligent-Unroll}.
   	 	&
   	 	\tabincell{l}{
   	 		(1) PageRank compiled by ICC.\\
   	 		(2) The method proposed by Peng Jiang\cite{jiang2018conflict}.\\
   	 		(3) The code generated by Intelligent-Unroll}.
   	 	 \\ \hline
   	     \tabincell{l}{ \textbf{Intel Xeon E5-2620 v3}\\
   	     (6 cores@2.4GHz,230.40 DP GFlops\\
   	      4 $\times$ DDR4,59 GB/s bandwidth).
         } &
     	  \tabincell{l}{
	      (1) The CSR-based SpMV compiled by ICC.\\
	     (2) The CSR-based SpMV with Intel MKL version 2019 Update 3.\\
    	 (3) CSR5-base SpMV\cite{liu2015csr5}.\\
	     (4) The code generated by Intelligent-Unroll}.
     	 &	
	      \tabincell{l}{
    	 	(1) PageRank compiled by ICC.\\
     		(2) The code generated by Intelligent-Unroll}.\\\hline
	\end{tabular} \label{tab-platform}
\end{table*}
We evaluate Intelligent-Unroll on two representable benchmarks, Sparse Matrix-Vector Multiply (SpMV) and PageRank. The code snippets of SpMV and PageRank are shown in Algorithm~\ref{alg-spmv} and Algorithm~\ref{alg-pagerank} respectively. We choose these two benchmarks due to their unique memory and calculation patterns. From Algorithm~\ref{alg-spmv}, we can see that in SpMV it always writes to the same memory location. Whereas for PageRank in Algorithm~\ref{alg-pagerank}, it exhibits a random memory write pattern. In addition, the calculation pattern of SpMV is represented by explicit reduction operations, whereas the reduction operations in PageRank are implicit. 

The experiment platform is an Intel Xeon Phi CPU (KNL) and an Intel Xeon CPU. The details of the platform and evaluation approach are shown in Table \ref {tab-platform}. The CPU machine is installed with 64-bit Ubuntu v16.04, whereas the KNL machine is installed with CentOS 7.4. The \textit{icc} v19.0.3 and LLVM v8.0.0 are installed on both machines. For SpMV, we compare to the implementations using CSR5~\cite{liu2015csr5} and MKL in addition to the default compiler optimization. For PageRank, we compare to the implementation using conflict-free method~\cite{jiang2018conflict} on KNL in addition to the default compiler optimization. We omit the results of conflict-free method on CPU since it does not support CPU architecture. The default compiler optimization of SpMV and PageRank uses \textit{icc} (-O3 -Xhost) that serves as our baseline. For each run, we execute the benchmark for 1,000 times, and measure the average execution time. Every experiment is evaluated for 10 times and the best result is reported. 

We select eight datasets from the University of Florida Sparse Matrix Collection to evaluate SpMV. The datasets include regular matrices such as \textit{Dense} and \textit{QCD}, as well as irregular matrices such as \textit{mip1} and \textit{Webbase-1M}. The datasets for evaluating PageRank are adopted from~\cite{jiang2018conflict}. The details of the evaluation datasets are shown in Table~\ref{tab-dataset}.

\subsection{Performance Opportunity Analysis}
\begin{table}
	\centering
	\scriptsize
	\caption{The datasets used by SpMV and PageRank.}
	\begin{tabular}{c|cccc}
		\hline\hline
		Benchmark 		 &Dataset       & row$\times$col & nnz & nnz/row \\\hline\hline
\multirow{8}{*}{SpMV}&Dense        &2K$\times$2K    &4.0M & 2K      \\
		    		 &FEM\_Ship    &141K$\times$141K&7.8M & 55     \\
		    		 &dc2      &117K$\times$117K&766K& 7     \\
		    		 &mip1         &66K$\times$66K  &10.4M&155     \\
		    		 &Webbase1M    &1M $\times$1M   &3.1M &3     \\
		    		 &Wind Tunnel &218K$\times$218K&11.6M&53      \\
		    		 &CirCuit      &171K$\times$171K&959K &5      \\
		    		 &QCD          &49K $\times$49K &1.9M &39     \\\hline\hline
\multirow{3}{*}{PageRank}&amazon0312&401K$\times$401K&3.2M&8K\\
					 &higgs-twitter &457K$\times$451K&15M &33K\\
					 &soc-pokec     &1.6M$\times$1.6M&31M &19.3    \\\hline
	\end{tabular} \label{tab-dataset}
\end{table}
\begin{table*}
	\centering
	\scriptsize
	\caption{The percentage of the \textit{gather}/\textit{scatter}/\textit{reduction} instructions that can be optimized by the \textit{load}/\textit{store} operation instructions for both SpMV and PageRank benchmarks across different datasets. The results are evaluated on CPU processor with vector length of 8.}
	\begin{tabular}{c|c|cccccccc|ccc}
		\hline\hline
		\multirow{2}{*}{ }&Benchmark& \multicolumn{8}{c|}{SpMV}&\multicolumn{3}{c}{PageRank}\\\hline
		&Detaset &Dense&FEM\_Ship&dc2&mip1&Webbase1M&Wind Tunnel&CirCuit &QCD &amazon0312&higgs-twitter&soc-pokec\\\hline
		\multirow{8}{*}{\textit{Gather}\&\&\textit{Scatter}}
		&L/S = 1 & 100\%& 15.1\% & 14.8\%&92.5\%& 5.6\%&61.7\%& 2.6\%&40.3\%
		&50.2\%&50.9\%&50.2\%\\
		&L/S = 2 & 0\%  & 84.9\% & 9.4\%& 1.5\%&53.0\%&37.8\%&22.5\% &45.8\%
		&1.3\%&0\%&0.5\%\\		
		&L/S = 3 & 0\%&  0\%    & 15.7\%& 1.2\%&18.1\%& 0.5\%&28.9\%&13.9\%
		&5.0\%&0\%&0.9\%\\
		&L/S = 4 & 0\%&  0\%    & 23.6\%& 1.3\%&11.0\%&   0\%&22.9\% &0\%
		&10.0\%&0.1\%&1.5\%\\
		&L/S = 5 & 0\%&  0\%    & 20.7\%& 1.7\%& 5.4\%&   0\%&14.6\%&0\%
		&12.1\%&0.2\%&2.4\%\\
		&L/S = 6 & 0\%&  0\%    & 10.1\%& 1.3\%& 3.0\%&   0\%&6.5\% &0\%
		&11.1\%&0.7\%&4.1\%\\
		&L/S = 7 & 0\%&  0\%    &  4.4\%& 0.1\%& 1.6\%&   0\%&1.6\%&0\%
		&7.2\%&4.2\%&9.0\%\\
		&L/S = 8 & 0\%&  0\%    &  1.3 \%& 0.4\%& 2.3\%&  0\%&0.4\%&0\%
		&3.1\%&44.8\%&31.4\%\\\hline
		\multirow{4}{*}{Reduction}
		&Op=0 &0\%  & 0\%   & 6.5\%& 0.6\%& 31.8\%& 1.0\%&18.4\%&2.6\%&
		92.0\%&100\%&100\%\\
		&Op = 1 &0\%  & 2.6\% & 5.8\%& 0.7\%& 32.5\%& 2.8\%&18.4\%&2.6\%
		&8.0\%&0\%&0\%\\
		&Op = 2 &0\%  & 4.7\& & 9.8\%& 1.3\%&  8.1\%& 3.7\%&32.5\%&5.1\%
		&0\%&0\%&0\%\\
		&Op = 3 &100\%& 92.7\%&77.9\%&97.4\%& 27.6\%& 92.5\%&30.7\%&89.7\%
		&0\%&0\%&0\%\\\hline\hline
		
	\end{tabular} \label{tab-dataset-distri}
\end{table*}

In Table~\ref{tab-dataset-distri}, we present the percentage of \textit{gather}/\textit{scatter}/\textit{reduction} instructions that can be replaced by \textit{load}/\textit{store}(\textit{L/S}) and \textit{vector} (\textit{Op}) instructions for the two benchmarks under different datasets. The second column in Table~\ref{tab-dataset-distri} indicates the number of \textit{load}/\textit{store}/\textit{vector} instructions that should be used to replace the original \textit{gather}/\textit{scatter}/\textit{reduction} instruction. We do not include the results of the \textit{scatter} instruction in SpMV, since they can be optimized by the statical analysis of compiler. The higher value of \textit{L/S} means the higher cost of replacing the \textit{gather}/\textit{scatter} instruction. Whereas $\textit{Op}=0$ means all \textit{reduction} instructions can be replaced with {vector} instructions, and $\textit{Op}=3$ means using the \textit{reduction} instruction supported by underlying architecture achieves better performance.


The SpMV running on the \textit{Dense} dataset illustrates a perfect case for instruction optimization in Table~\ref{tab-dataset-distri}, where each of its \textit{gather} instructions can be replaced with only one \textit{load} instruction. In addition, we can optimize the reduction operations ($\textit{Op}=3$) with the reduction instruction provided by underlying architectures. There are also some cases where there is hardly any performance opportunity with Intelligent-Unroll, such as \textit{Webbase-1M} and textit{CirCuit} whose $L/S=1$ is less than 6\%. Compared to SpMV, the datasets of PageRank are more irregular. With $L/S=1$, the percentage of replaceable instructions is less than 51\%. And even with $L/S=8$, the percentage is no more than 44.8\% (e.g., \textit{higgs-twitter} dataset). 

\subsection{PageRank}
\begin{algorithm}[t]
	\caption{The PageRank defined in Intelligent-Unroll}
	\label{alg-pagerank-def}
	\begin{algorithmic}[1] \footnotesize
		\State $input: $ \label{alg-pagerank-def-input-begin}
		\State $\ \ \ $int * $n1$, int * $n2$, double * $rank$, double * $nneighbor$\label{alg-pagerank-def-input-end}
		\State $output:$\label{alg-pagerank-def-output-begin}
		\State $\ \ \ $double *$ sum$\label{alg-pagerank-def-output-end}
		\State $lambda\ i\ :$\label{alg-pagerank-def-expr-begin}
		\State $\ \ \ sum[ n2[i] ] \gets sum[n2[i]]\  +\  rank[n1[i]] \times nneighbor[n1[i]]$\label{alg-pagerank-def-expr-end}
		
	\end{algorithmic}
\end{algorithm}
The code snippet of PageRank shown in Algorithm~\ref{alg-pagerank} can be defined using Intelligent-Unroll as Algorithm~\ref{alg-pagerank-def}. The keyword \textit{input} (line~\ref{alg-pagerank-def-input-begin}-\ref{alg-pagerank-def-input-end}) and \textit{output} (line~\ref{alg-pagerank-def-output-begin}-\ref{alg-pagerank-def-output-end}) define the inputs and outputs of the PageRank algorithm respectively. The lambda expression specifies the calculation details (line~\ref{alg-pagerank-def-expr-begin}-\ref{alg-pagerank-def-expr-end}). Based on Algorithm~\ref{alg-pagerank-def}, we can generate an implementation from Intelligent-Unroll for the PageRank algorithm. 
\begin{table*}
	\centering
	\scriptsize
	\caption{The performance of PageRank across different datasets on KNL and CPU.}
\begin{tabular}{ccc}
	
	\includegraphics[scale=0.14]{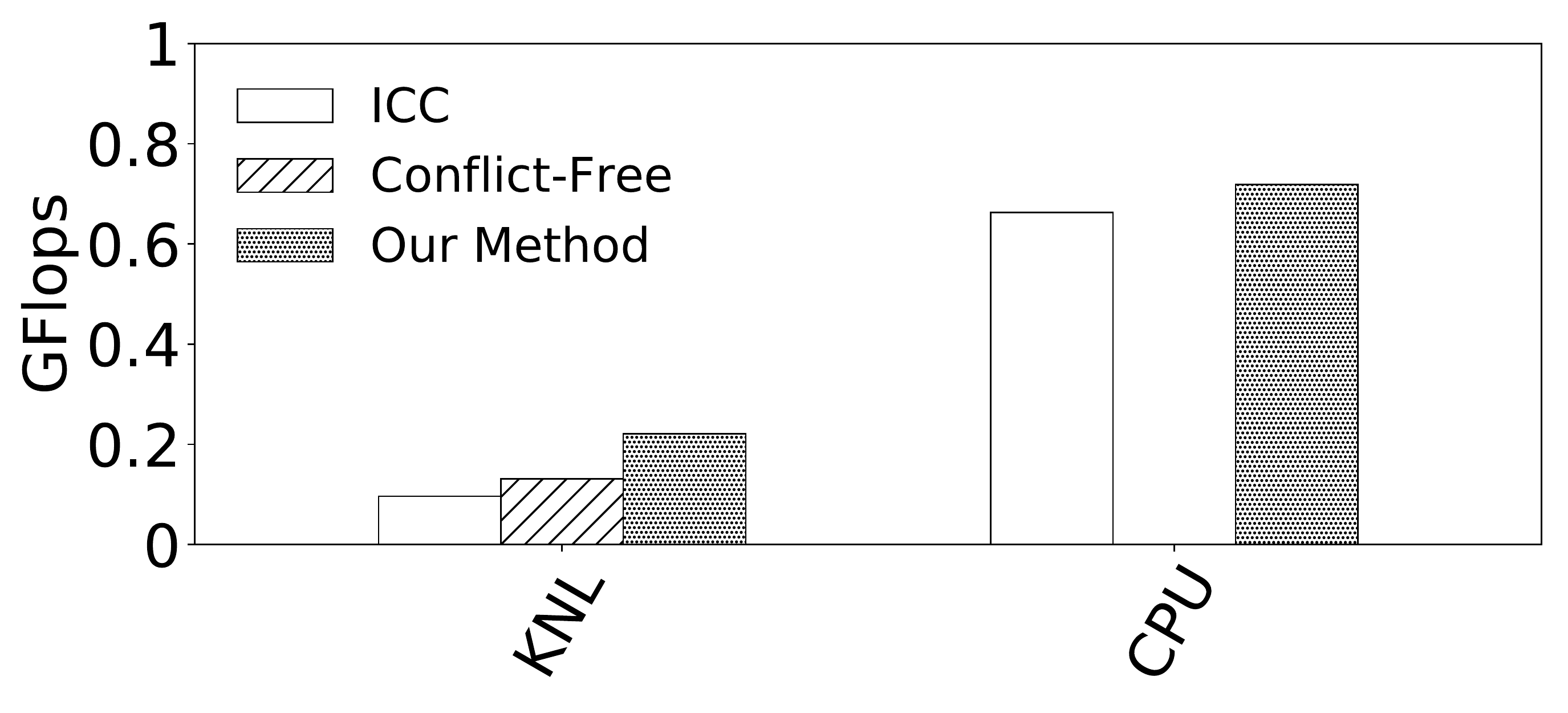} & 
	\includegraphics[scale=0.14]{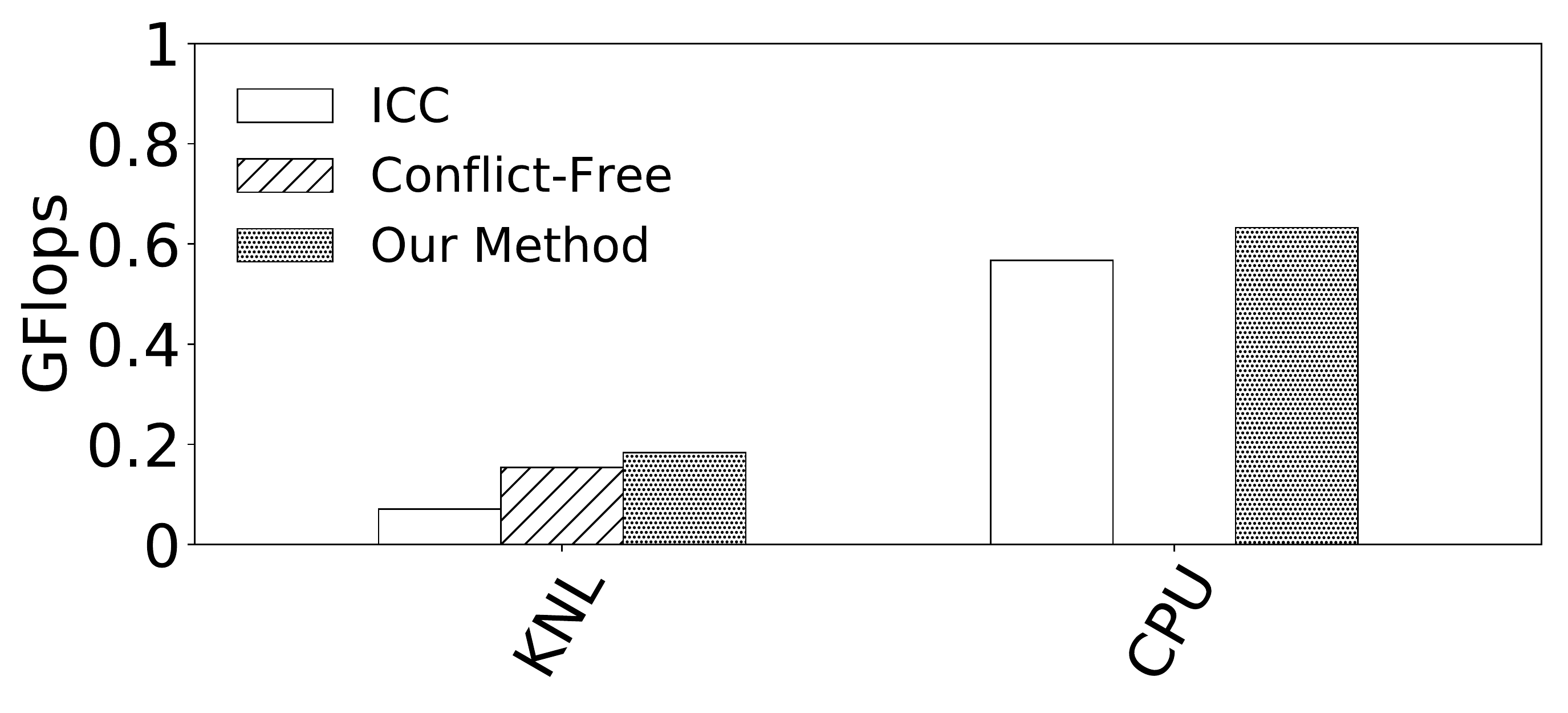} &
	\includegraphics[scale=0.14]{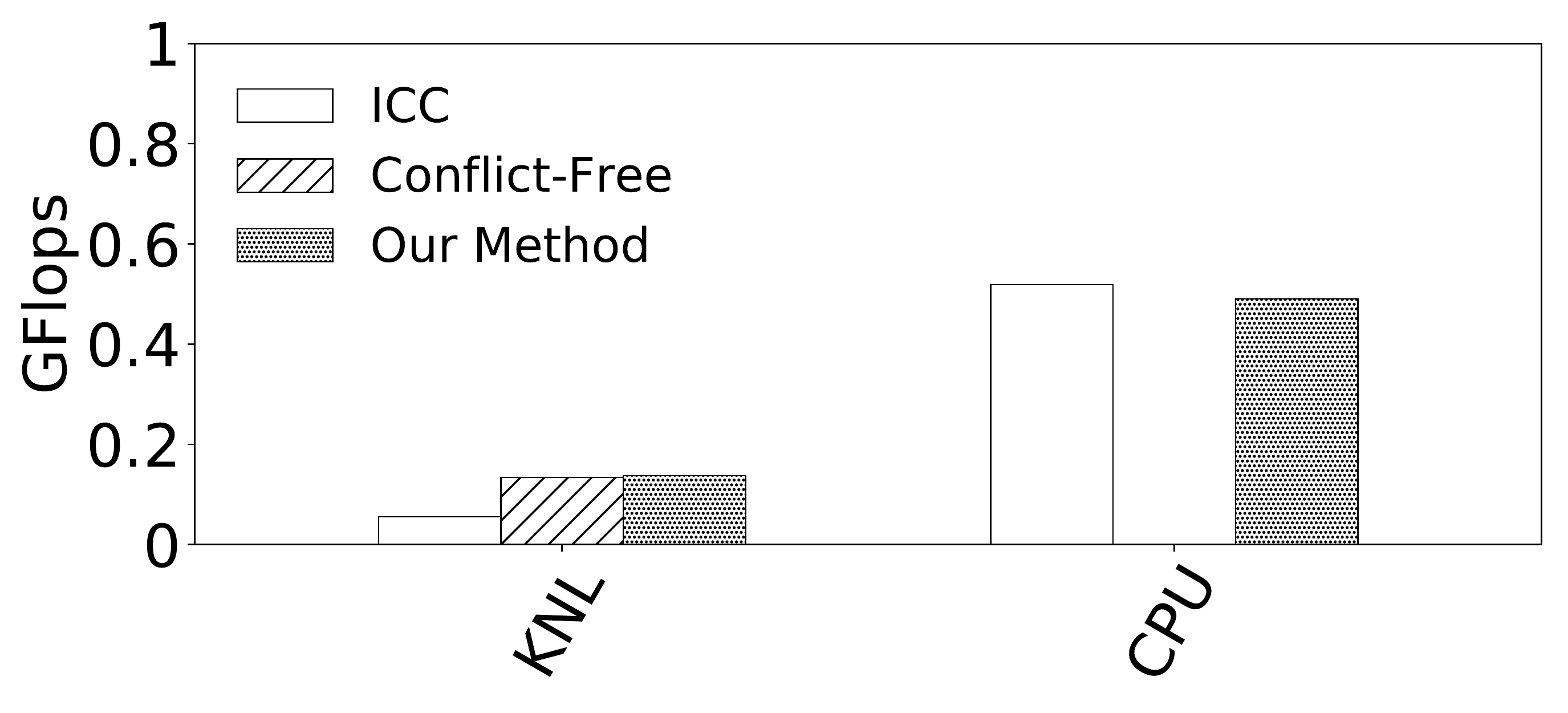} \\
	amazon0312 & higgs-twitter & soc-pokec\\
	
\end{tabular} \label{fig-pagerank-performance}
\end{table*}

Table~\ref{fig-pagerank-performance} shows the performance comparison of PageRank implemented using the methods of Intelligent-Unroll, conflict-free and default compiler optimization on KNL and CPU. We can see that the implementation optimized by our method achieves better performance across almost all datasets on both KNL and CPU. Our method improves the performance of PageRank by 4.8\% on average (11.6\% on maximum) compared to the baseline on CPU, and by 30.2\% and 146.0\% on average (68.5\% and 158.8\% on maximum) compared to the conflict-free and baseline methods respectively on KNL.

On KNL, with \textit{higgs-twitter} and \textit{soc-pokec} datasets, our method achieves similar performance to conflict-free method, both of which is better than the baseline. This is because \textit{icc} cannot optimize PageRank due to the potential write conflicts. However, on \textit{amazon0312} dataset, the performance of our method outperforms the rest by a large margin. This is because the percentage of instructions with $Op=1$ on \textit{amazon0312} dataset is more than 8\%, which are  also randomly distributed during the execution. Such random distribution degrades the effectiveness of branch prediction in the conflict-free method. However, in Intelligent-Unroll, the code is directly generated for each branch condition without predicting during runtime. Therefore, our method outperforms the conflict-free method on \textit{amazon0312} dataset.


On CPU, with \textit{amazon} and \textit{higgs-twitter} datasets, the performance using Intelligent-Unroll is better than the default compiler optimization. However on \textit{soc-pokee} dataset, the code generated by Intelligent-Unroll is slower than the code optimized by \textit{icc}. This is because the vector length is quite limited (e.g., 8 in single precision) on CPU that outsets the performance benefit when replacing the reduction instructions with vector instructions.   

The reason why our method achieves better performance than the conflict-free method on KNL can be attributed to two folds: \textit{1)} our method generates the code for each data access and instruction pattern of PageRank. Therefore, it avoids pattern prediction during runtime and thus improves performance; \textit{2)} in addition to use SIMD instruction, our method also replaces the \textit{gather}/\textit{scatter} instructions with \textit{load}/\textit{store} instructions. As shown in Table~\ref{tab-dataset-distri}, the percentage of instructions with $L/S=1$ across all datasets is larger than 50\%, which presents significant opportunity for performance improvement of PageRank.

\subsection{SpMV}
\begin{algorithm}[t]
	\caption{The SpMV defined in Intelligent-Unroll}
	\label{alg-spmv-def}
	\begin{algorithmic}[1] \footnotesize
	\State $input: $ \label{alg-spmv-def-para-begin}
	\State $\ \ \ $int * $row\_ptr$,int * $col\_ptr$,double * $x$,double * $value$
	\State $output:$
	\State $\ \ \ $double *$ y$\label{alg-spmv-def-para-end}
	\State $lambda\ i\ :$\label{alg-spmv-def-expr-begin}
	\State $\ \ \ y[ row\_ptr[i] ] \gets y[row\_ptr[i]]\  +\ value[i] \times x[column\_ptr[i]]$\label{alg-spmv-def-expr-end}
	\end{algorithmic}
\end{algorithm}
The baseline SpMV implementation uses CSR format, because it decreases memory usage and provides more opportunity for compiler optimization. However in Intelligent-Unroll, we use COO instead of CSR which fits well with our optimization method. Algorithm~\ref{alg-spmv-def} defines SpMV using Intelligent-Unroll (line~\ref{alg-spmv-def-expr-begin}-\ref{alg-spmv-def-expr-end}). We can see that the definition using Intelligent-Unroll is more concise than the original definition in Algorithm~\ref{alg-spmv}. Intelligent-Unroll automatically optimizes the data access and instruction instead of relying on manual optimization. After defining the calculation, users only need to specify the \textit{input} and \textit{output} (line~\ref{alg-spmv-def-para-begin}-\ref{alg-spmv-def-para-end}) in Intelligent-Unroll. Table~\ref{fig-spmv-performance} shows the performance comparison among the methods using default compiler optimization, MKL, CSR5 and our method on both CPU and KNL. 

\begin{table*}
	\centering
	\scriptsize
	\caption{The performance of SpMV across different datasets on KNL and CPU.}
\begin{tabular}{cccc}
	
	\includegraphics[scale=0.14]{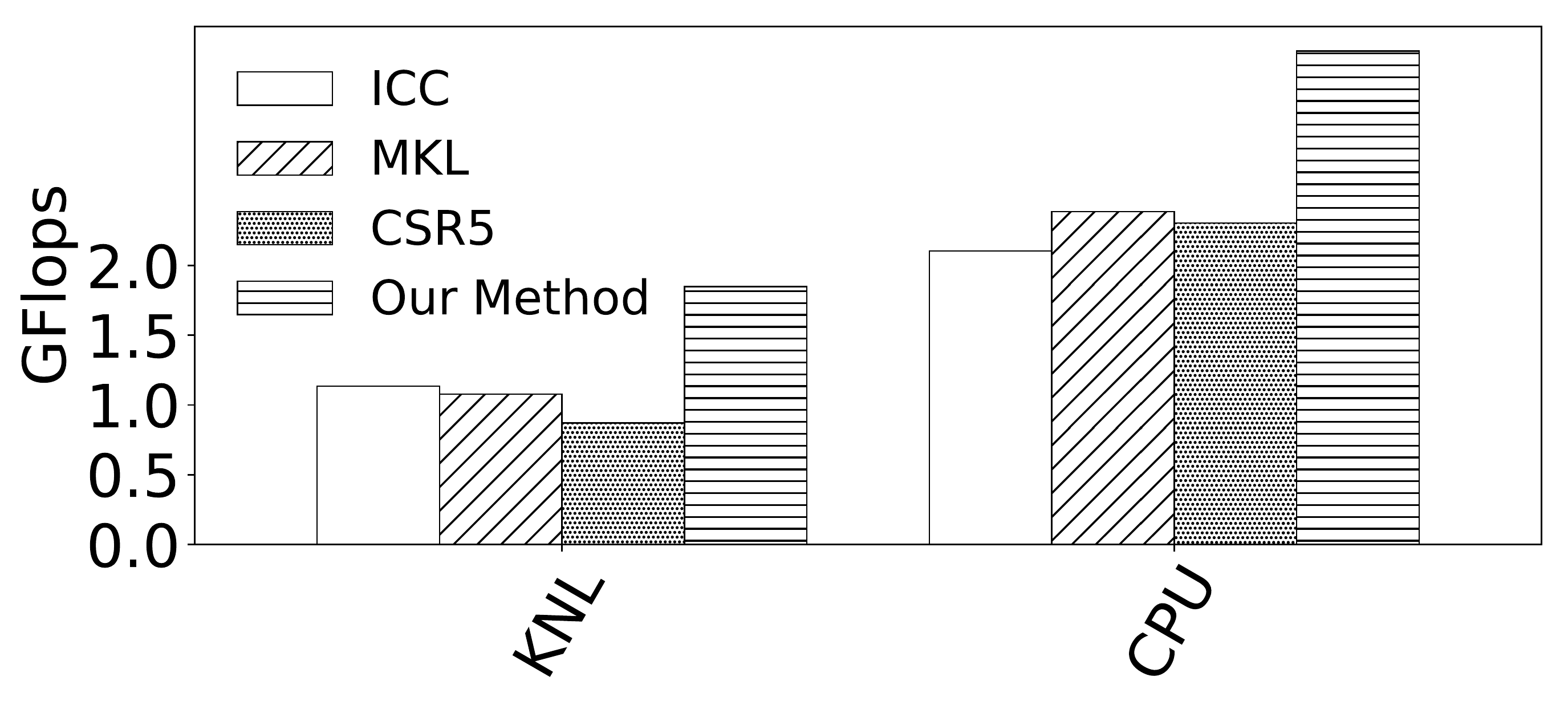} & 
	\includegraphics[scale=0.14]{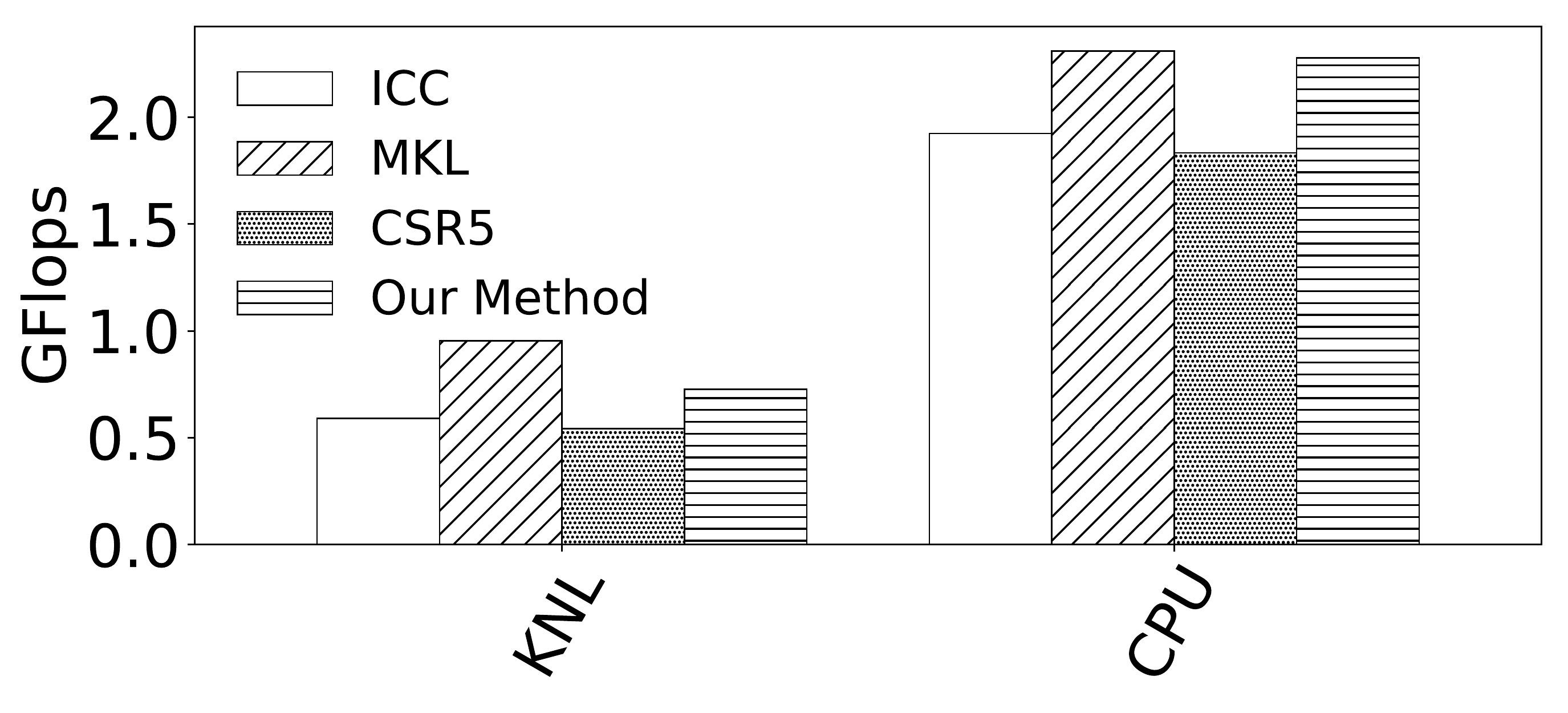} &
	\includegraphics[scale=0.14]{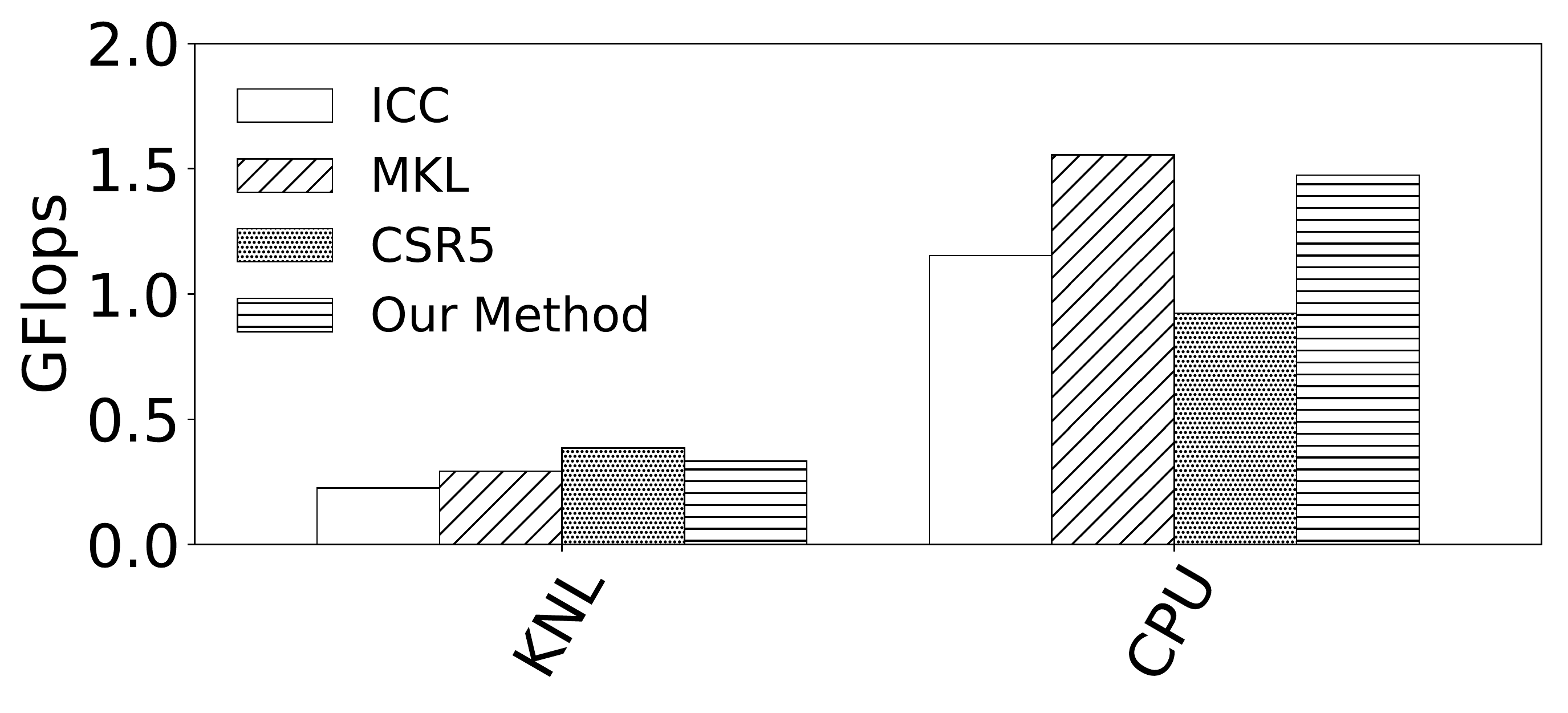} &
	\includegraphics[scale=0.14]{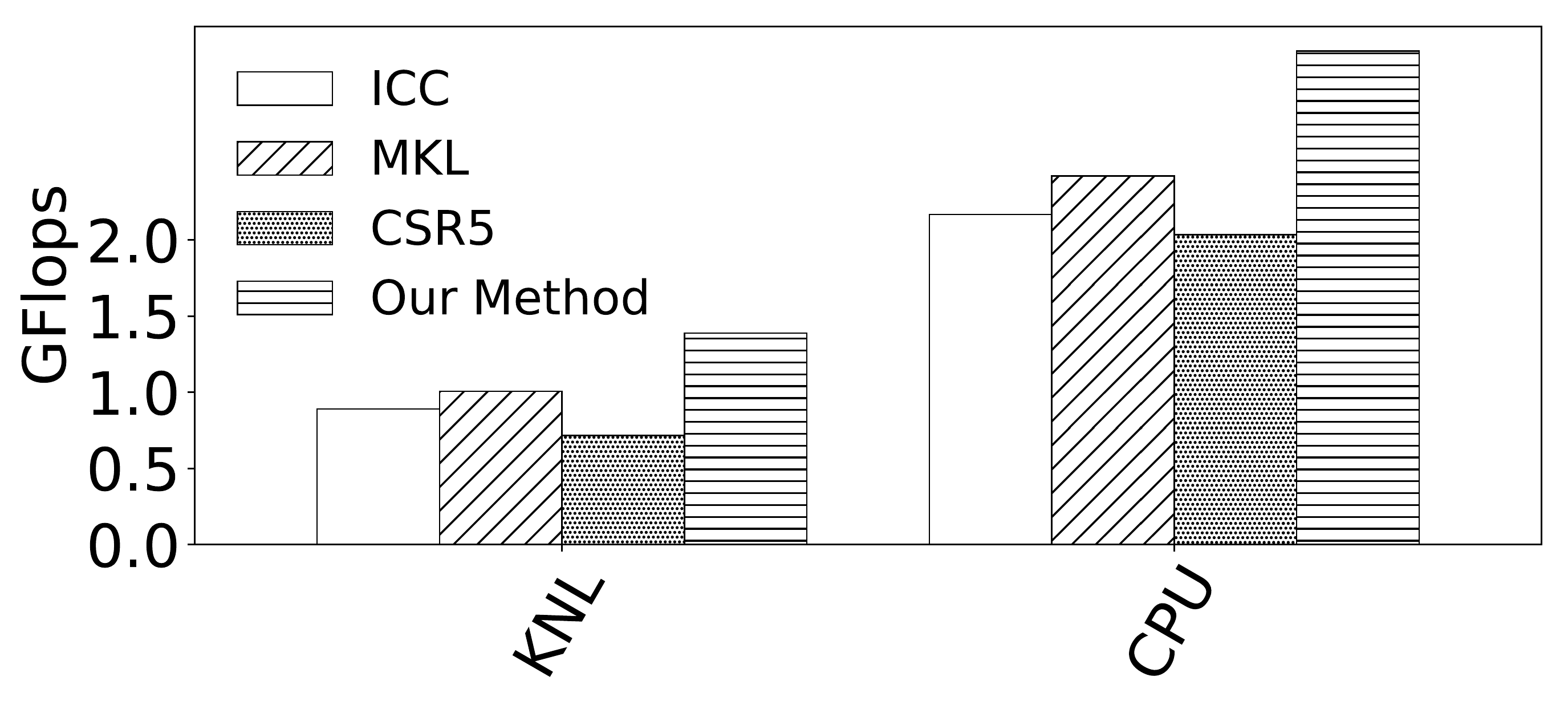} \\
	Dense & FEM\_Ship &dc2&mip1\\
	\includegraphics[scale=0.14]{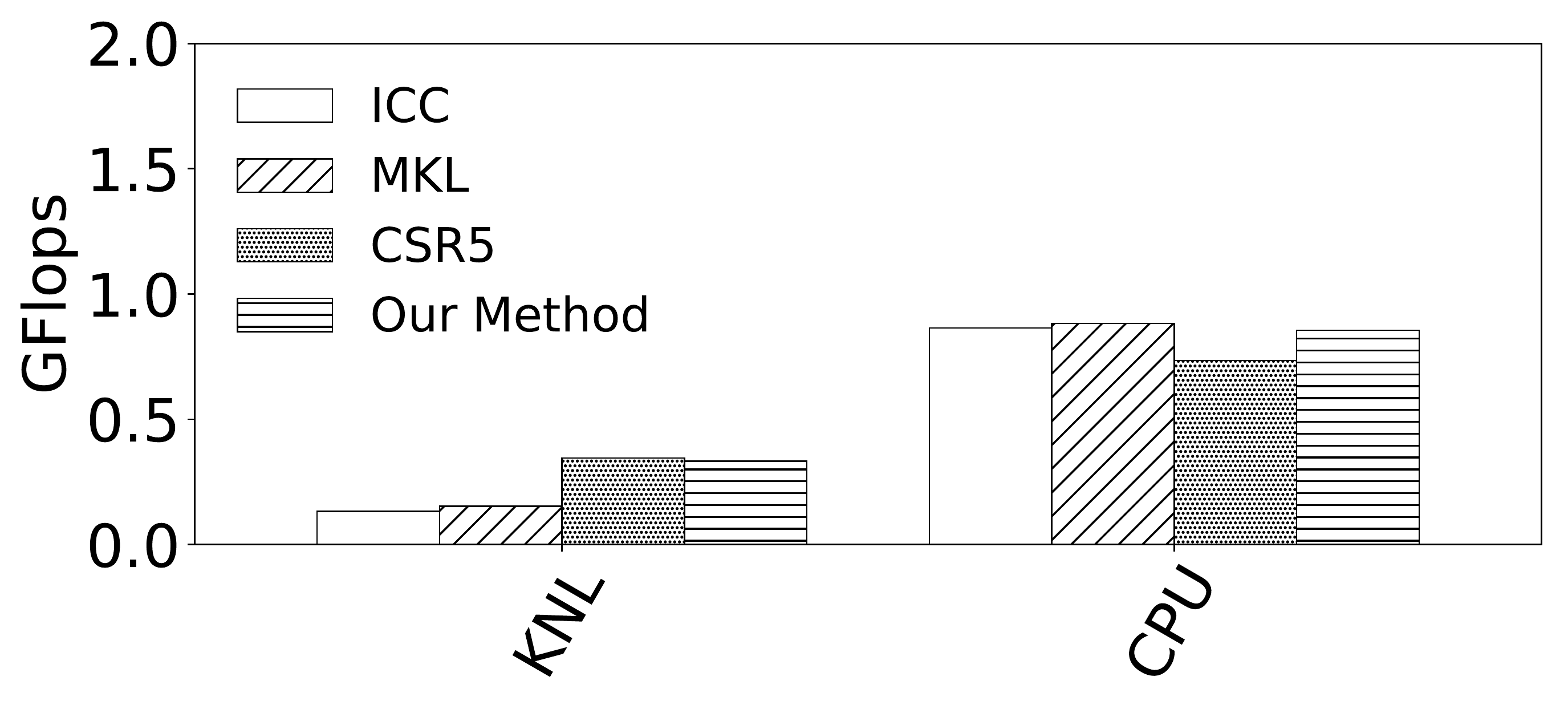} & 
	\includegraphics[scale=0.14]{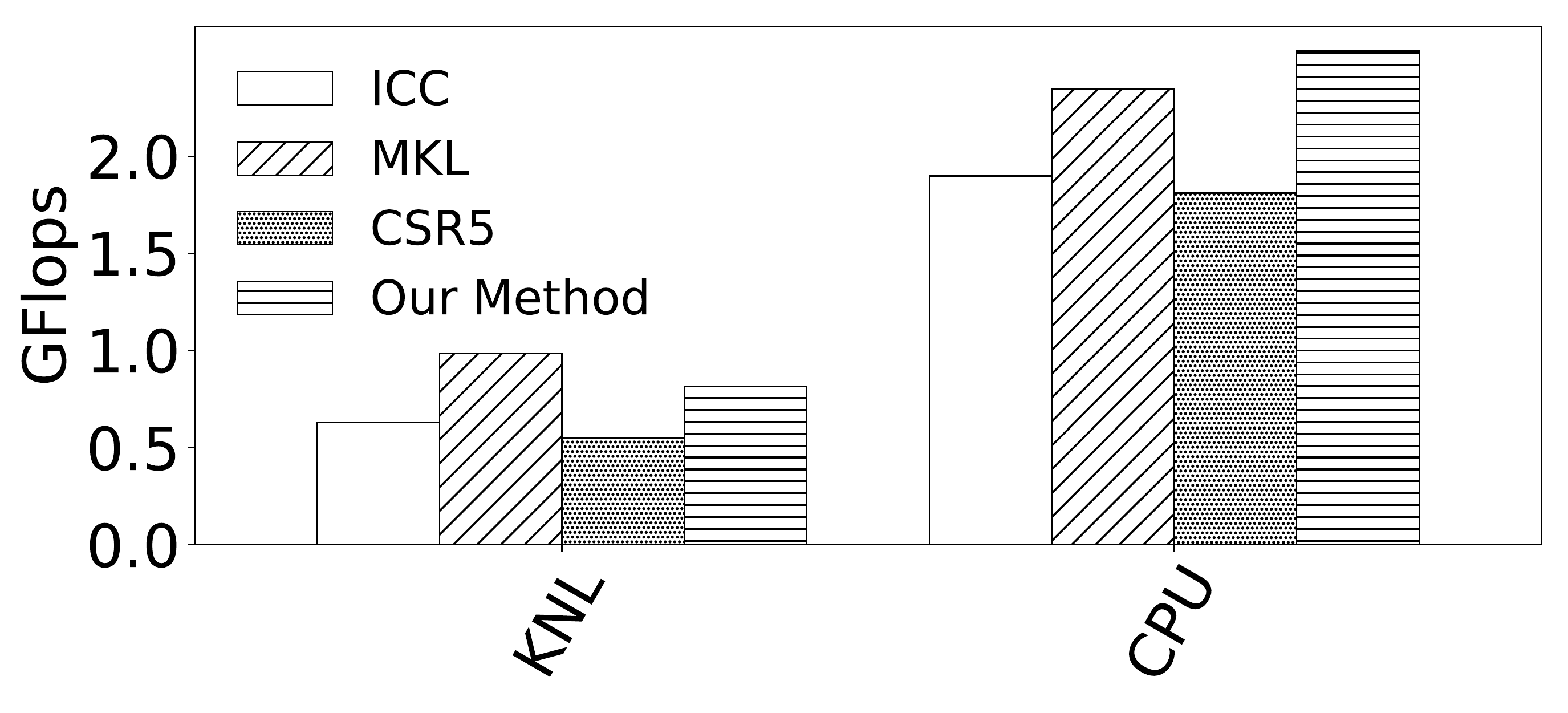} &
	\includegraphics[scale=0.14]{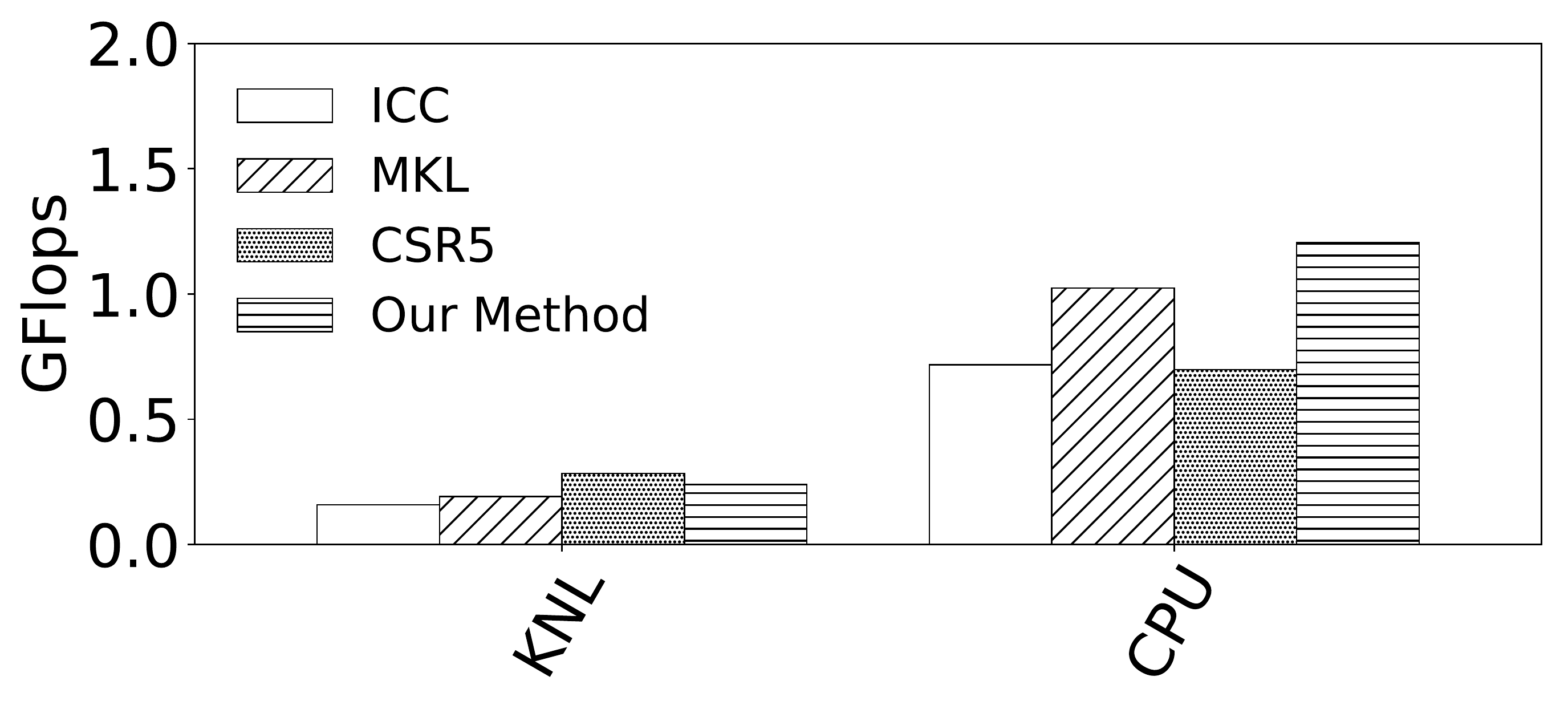} &
	\includegraphics[scale=0.14]{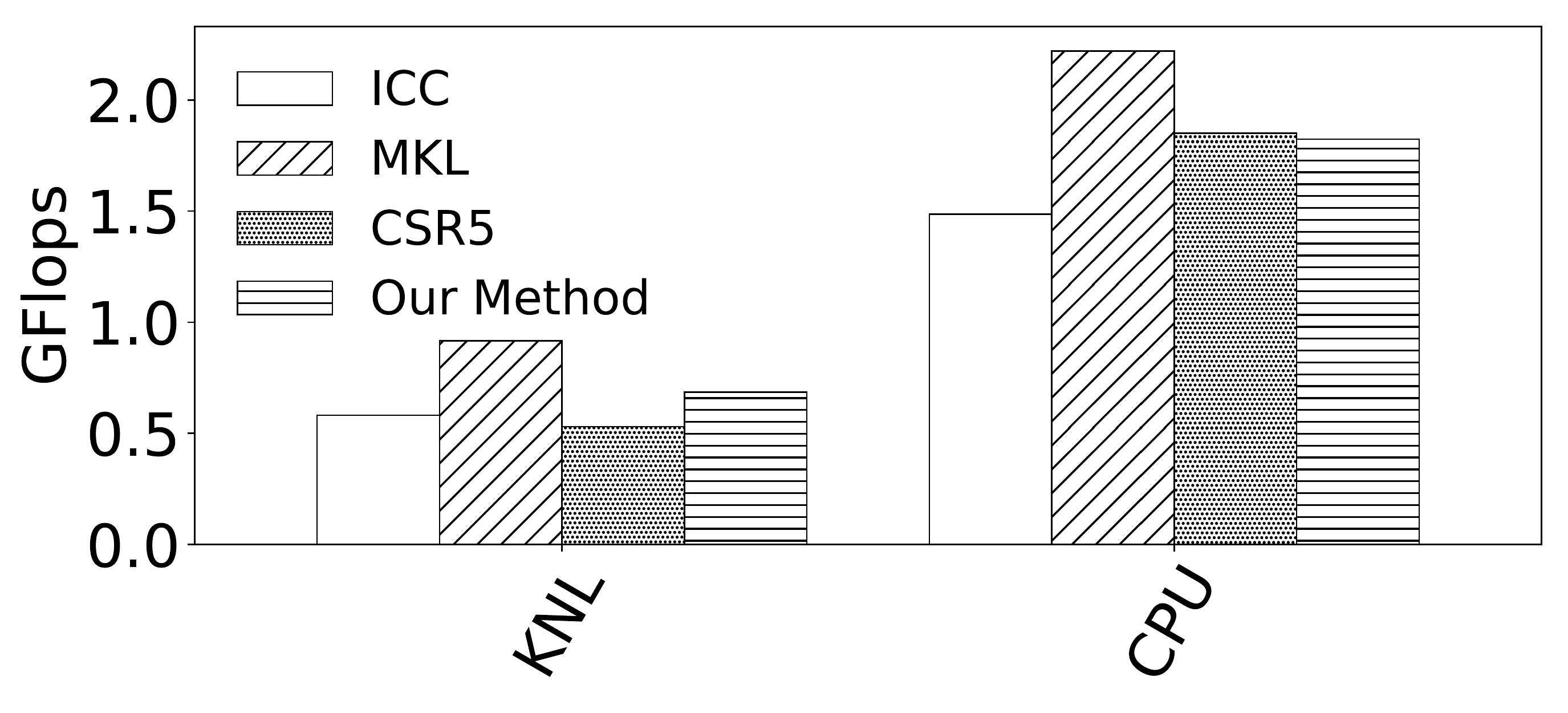}\\
	Webbase1M & Wind Tunnel & CirCuit &QCD\\
	
\end{tabular} \label{fig-spmv-performance}
\end{table*}


On KNL, our method achieves best performance on \textit{Dense} and \textit{mip1} datasets, whereas on \textit{FEM\_Ship}, \textit{Wind Tunnel} and \textit{QCD} datasets, the MKL implementations achieve best performance. The CSR5 implementations also achieve best performance on \textit{dc2} and \textit{CirCuit} datasets. The reason why our method achieves better performance than other methods is similar to the PageRank benchmark. This is because our method is able to avoid branch prediction during runtime and improve the memory accesses with \textit{load}/\textit{store} instructions. However, on the datasets where the MKL implementation is better, the reason can be attributed to 
the split of the writes to the same memory location from different calculation patterns in our method, which increases the \textit{load}/\textit{write} instructions to the output vector \textit{y}. On datasets where CSR5 achieves best performance, it is because the data structure of input matrices is friendly to CSR5 format and corresponding calculation pattern, which has not been integrated in the \textit{optimization pass} of \textit{Intelligent-Unroll} yet.

On CPU, our method achieves the best performance on \textit{Dense}, \textit{mip1}, \textit{Wind Tunnel} and \textit{CirCuit} datasets. On the datasets where our method fails to achieve the best performance, the reason can be attributed to the limited vector length on CPU that diminishes the advantage of Intelligent-Unroll by avoiding the branch prediction during runtime due to the small number of conditions.
%
In sum, compared to baseline, MKL and CSR5, our method improves the performance of SpMV by 54.8\%, 24.9\% and 35.7\% on average (151.0\%, 116.9\%, 112.0\% on maximum) respectively on KNL, whereas by 35.9\%, 10.1\% and 40.5\% on average (68.2\%, 48.3\% and 72.5\% on maximum) on CPU. 

\section{Related Work}
\label{sec-relatedWork}

\textbf{Designing efficient sparse data formats - }Many sparse data formats are proposed targeting different sparsity patterns as well as the architecture diversity. For instance, block-based formats are widely adopted due to the cache-friendly design~\cite{ ashari2014efficient,  buluc2011reduced,bulucc2009parallel}. CSR5~\cite{liu2015csr5} and CVR~\cite{xie2018cvr} proposed new sparse data formats for SpMV, which focus on optimizing the instruction parallelism and load balance. Liu et al.~\cite{liu2013efficient} proposed ELLPACK to accelerate SpMV kernel on Intel KNL processor. Choi et al.~\cite{choi2010model} proposed to use small sub-blocks, each of which is represented as a dense matrix, to optimize SpMV on GPUs.

\textbf{Improving the temporal and spatial data reuse - }Since there are many sparse data formats available, determining the appropriate sparse format for the irregular application is not trivial. Friese et al.~\cite{daniel2004multiparameter} and Xie et al.~\cite{xie2019ia} proposed different performance models to determine the optimal sparse data format. In essence, their works optimize the irregular applications by improving the temporal reuse and spatial data reuse with the appropriate sparse format. 
There are also many works exploring optimization works on distributed memory architectures\cite{das1994communication,basumallik2006optimizing}.
Several loop unrolling strategies~\cite{stephenson2005predicting,kisuki2000combined,mellor2004optimizing} are proposed in literature. However, these works mainly focused on selecting optimal tile size and unroll factor when unrolling the loop, and failed to exploit the performance opportunity by optimizing the instructions.

\textbf{Optimizing parallelization strategies - }Different parallelization strategies were proposed when optimizing the irregular applications on specific architectures~\cite{kulkarni2008scheduling,kulkarni2009lonestar}. 
Jiang et al.~\cite{jiang2018conflict} optimized the irregular applications by parallelizing the computation using the powerful SIMD units. Buobo et al.~\cite{buono2015optimizing} proposed optimizations of sparse linear algebra tailored for large-scale graph analytics. Millind et al~\cite{kulkarni2009much} proposed a tool called ParaMeter to profile parallelism information of irregular programs.

\section{Conclusion}
\label{sec-conclusion}
In this paper, we address the limitation of traditional compilers that is unable to exploit the performance opportunity for optimizing irregular applications due to its static analysis. We propose our solution of Intelligent-Unroll that identifies the regular patterns within irregular applications, and automatically optimizes the data access and instruction for generating more efficient code. The experiment results with representative benchmarks on both CPU and KNL processors demonstrate the effectiveness of our approach in optimizing the irregular applications for better performance compared to the-state-of-the-art implementations.
\bibliography{subsection/bibfile}







\end{document}